\documentclass[useAMS,usenatbib,usegraphicx,onecolumn]{mn2e}

\def\lapp{\ \lower 3pt\hbox{${\buildrel < \over \sim}$}\ }
\def\gapp{\ \lower 3pt\hbox{${\buildrel > \over \sim}$}\ }
\def\gl{\ \lower 3pt\hbox{${\buildrel > \over {\scriptscriptstyle <}}$}\ }
\def\lg{\ \lower 3pt\hbox{${\buildrel < \over {\scriptscriptstyle >}}$}\ }
\newcommand{\be}{\begin{equation}}
\newcommand{\ee}{\end{equation}}
\newcommand{\bea}{\begin{eqnarray}}
\newcommand{\eea}{\end{eqnarray}}
\newcommand{\bean}{\begin{eqnarray*}}
\newcommand{\eean}{\end{eqnarray*}}

\newcommand{\next}{\nonumber\\}
\newcommand{\rn}[1]{(\ref{#1})}
\newcommand{\ff}[2]{{\textstyle \frac{#1}{#2}}}
\newcommand{\hand}{\hspace{0.5cm}{\rm and}\hspace{0.5cm}}
\newcommand{\dg}{{\rm o}}
\newcommand{\bOmega}{{\mbox{\boldmath$\Omega$}}}
\newcommand{\sspace}{\phantom{.}\vspace{0.0mm} \noindent}
\newcommand{\hh}{\hspace{0.5cm}}
\newcommand{\ben}{\begin{enumerate}}
\newcommand{\een}{\end{enumerate}}

\title[Tidal evolution of inclined systems]
{The determination of planetary structure in tidally relaxed inclined systems}
\author[Rosemary A. Mardling]{Rosemary A. Mardling$^{1}$\thanks{E-mail:
mardling@sci.monash.edu.au}\\
$^{1}$School of Mathematical Sciences, Monash University, Victoria, 3800, Australia}
\begin{document}

\date{Accepted ... Received ...; in original form ...}

\pagerange{\pageref{firstpage}--\pageref{lastpage}} \pubyear{2010}

\maketitle

\label{firstpage}

\begin{abstract}

The recent discovery of a transiting short-period planet on a slightly non-circular orbit
with a massive highly eccentric companion orbiting the star HAT-P-13 offers the possibility of probing the structure of 
the short-period planet. The ability to do this relies on the system being
in a quasi-equilibrium state in the sense that the eccentricities are constant on 
the usual secular timescale (typically, a few thousand years), and decay on a timescale which is
much longer than the age of the system. 
Since the equilibrium eccentricity is effectively a function only of observable system parameters
and the unknown Love number of the short-period planet, the latter can be determined with
accurate measurements of the planet's eccentricity and radius.

However, this analysis relies on the assumption that
the system is coplanar, a situation which seems unlikely given the high eccentricity of the outer planet.
Here we generalize our recent analysis of this fixed-point phenomenon to mutually inclined systems
in which the outer body dominates the total angular momentum,
and show that (1) the fixed point of coplanar systems is replaced by a {\it limit cycle} in $e_b-\eta$ space, where $e_b$ is the
eccentricity of the inner planet and
$\eta$ is the angle between the periapse lines, with the average value of $e_b$, $e_b^{(av)}$, decreasing and its
amplitude of variation increasing with increasing mutual inclination.
This behaviour significantly reduces the ability to unambiguously determine the Love number
of the short-period planet if the mutual inclination is higher than around 10$^\dg$.
(2) We show that for $Q$-values less than $10^6$, the HAT-P-13 
system cannot have a mutual inclination between 54 and 126 degrees
because Kozai oscillations coupled with tidal dissipation
would act to quickly move the inclination outside this range,
and (3) that the behaviour of retrograde systems is the mirror image of that for prograde systems
in the sense that (almost) identical limit cycles exist for a given mutual inclination and $\pi$ minus this value.
(4) We derive a relationship between $e_b^{(av)}$, the equilibrium radius of the short-period planet,
its $Q$-value and its core mass,
and show that given current estimates of $e_b$ and the planet radius, 
as well as the lower bound placed on the $Q$-value by the decay rate of $e_b^{(av)}$,
the HAT-P-13 system is likely to be close to prograde coplanar,
or have a mutual inclination between $130^\dg$ and $135^\dg$.
Lower rather than higher core masses are favoured.
(5) An expression for the timescale for decay of the mutual inclination is derived, revealing that it evolves
towards a non-zero value as long as $e_b>0$ on a timescale which is much longer than the age of the system.
(6) We conclude with a scattering scenario for the origin of the HAT-P-13 system and show that
almost identical initial conditions can result in significantly different outer planet eccentricities,
stellar obliquities and planet radii.
The implications for 
systems with high stellar obliquities such as HAT-P-7 and WASP-17
are briefly discussed.

\end{abstract}


\section{Introduction}

Transiting systems offer the opportunity to determine a wide variety of system parameters,
including the mass of the transiting planet, its orbital eccentricity, the inclination of its orbit to the line of sight, its radius
and hence mean density, and the sky projection of
the angle between its orbit normal and the stellar spin axis. Many other system parameters
are potentially measurable \citep{winn0}, one of them being the Love number of the transiting planet 
if the system parameters are favourable \citep{wu}. 
The recent discovery of the HAT-P-13 system \citep{bakos} provides us with such a system as was recently 
pointed out by \citet{batygin}.
A reliable estimate of a planet's Love number in turn allows one to say something
about the presence or otherwise of a planetary core, and hence about the formation mode; in the 
former case the planet is likely to have been formed via core accretion of solid material and subsequent accretion
of a massive atmosphere \citep{pollack}, while the absence of a core supports the gravitational collapse hypothesis
of giant planet formation \citep{boss}.

The HAT-P-13 system consists of a 0.85 Jupiter-mass planet (planet b) in an almost circular 
2.9 day orbit about a 1.2 solar-mass star,
and a companion with a minimum mass of $15.2 M_J$ in a 428 day highly eccentric orbit (planet c).
Table~\ref{planet}
\begin{table*}
\label{planet}
 \centering
 \begin{minipage}{140mm}
  \caption{Parameters of the HAT-P-13 system}
\begin{tabular}{lr}
\hline
HAT-P-13 &\\
\hh $m_*$ ($M_\odot$) & $1.22^{+0.05}_{-0.10}$ \\
\hh $R_*$ ($R_\odot$) & $1.56\pm0.08$ \\
\hh age (Gyr) & $5.0^{+2.5}_{-0.8}$\\
\hh $k_*$ & 0.03\\
\hh $r_{g*}/R_*$ & 0.076\\
\hh $\gamma_*^{GR}$ & 2.04\\
\hh $\gamma_*^{tide}$ & 0.04\\
\hh $\gamma_*^{spin}$ [$P_{spin}=P_\odot$] & 0.06\\
&\\
HAT-P-13b &\\
\hh $m_b$ ($M_J$) & $0.851^{+0.029}_{-0.046}$ \\
\hh $a_b$ (AU) & $0.0426^{+0.0006}_{-0.0012}$ \\
\hh $P_b$ (days) & $2.916260\pm0.000010$ \\
\hh $e_b$ & $0.021\pm 0.009$ \\
\hh $\omega_b$ & $181\pm 46^{\rm o}$ \\
\hh $R_b$ ($R_J$) & $1.280\pm 0.079$ \\
\hh $i_{los}$ & $83.4\pm 0.6^\dg$\\
\hh $k_b$ & 0.3\\
\hh $r_{gb}/R_b$ & 0.26\\
\hh $\gamma_b^{tide}$ & 4.83\\
\hh $\gamma_b^{spin}$ & 0.32\\
\hh $\tau_{circ}$ & $40\,(Q_b/10^5)$ Myr \\
\hh $\tau_a$ & $79\,(Q_b/10^5)$ Gyr \\
\hh $\tau_e$ & $15\,(Q_b/10^5)$ Gyr \\
\hh $\tau_c$ & $9200\,(Q_b/10^5)$ Gyr \\
\hh $\tau_i$ & $80\,(Q_b/10^5)$ Gyr \\&\\
HAT-P-13c &\\
\hh $m_c \sin i_{los}\equiv m_c^{min}$ ($M_J$) & $15.2\pm 1.0$ \\
\hh $a_c$ (AU) & $1.186^{+0.018}_{-0.033}$ \\
\hh $P_c$ (days) & $428.5\pm 3.0$ \\
\hh $e_c$ & $0.691\pm 0.018$ \\
\hh $\omega_c$ & $176.7\pm 0.5^{\rm o}$\\
\hline
\end{tabular}
\end{minipage}
\end{table*}
lists the relevant parameters of the system, with data taken from \citet{bakos} who performed
a simultaneous fit of HATNet and KeplerCam photometric data and Keck spectroscopic data,
a process they refer to as ``global'' modelling.
Here $P_b$ and $P_c$ are orbital periods, $i_{los}$ is the inclination of the orbit normal of planet $c$ to the line of sight
and $\omega_b$ and $\omega_c$ are periapse arguments.
Note in particular the non-zero estimate of $0.021\pm 0.009$ for the inner planet's eccentricity. 
The main aim of this paper is to explore the extent to which information can be gleaned from such a non-zero measurement,
whether or not we know the mutual inclination of the orbits of the two planets.

The ability to determine the Love number of HAT-P-13b relies on the system being
in a quasi-equilibrium state in the sense that after an
initial transient phase of rapid tidal evolution,
the eccentricities change on a timescale much longer than the age of the system and
the apsidal lines of the inner and outer planets are aligned or anti-aligned.
Moreover, the mass and eccentricity of HAT-P-13c ensure that the quasi-equilibrium eccentricity of the inner planet
is significant and measurable.
Some of the theory for this is developed in \citet{wu} for the HD83443 system,
while a general theory is presented in \citet{puffball}.
Both of these studies, however, are for the coplanar case only.
Using the HAT-P-13 system as illustration, 
we generalize the theory to non-coplanar systems in which the outer body dominates the total angular momentum
and show that the quasi-relaxed state no longer corresponds to a fixed point in $e_b-\eta$ space, where 
$\eta=\varpi_b-\varpi_c$ with $\varpi_b$ and $\varpi_c$ the longitudes of periastron,
but rather corresponds to a {\it limit cycle} in this space (see, for example, Jordan \& Smith 1999), 
with the average value of $e_b$ decreasing and its
amplitude of variation increasing with increasing mutual inclination.
As in the coplanar case, the relaxation timescale is around three times the circularization timescale.
However, {\it unlike} the coplanar case, the rate of change of the {\it argument} of periastron of the inner planet
plays an important role, with the limit cycle frequency equal to twice this quantity.
In fact it is the appearance of this additional frequency which prevents the system from evolving to
a fixed point, with terms which depend on it effectively acting as external forcing terms.
Note that our analysis takes into account the fact that the actual mass of the outer planet increases with
increasing mutual inclination, given the observed minimum mass determined via radial velocity measurements \citep{bakos}.

The plan for this paper is as follows.
Section 2 reviews the theory for the long-term tidal evolution of coplanar systems, Section 3
generalizes this to mutually inclined systems, including limit cycle behaviour of prograde systems (Section 3.1),
the Kozai regime (Section~\ref{high}),
limit-cycle behaviour of retrograde systems (Section~\ref{retrograd}),
the relationship between $e_b^{(av)}$, the equilibrium radius of planet b, its $Q$-value and its
core mass (Section~\ref{seceReq}),
and the timescale for decay of the mutual inclination in a relaxed system (Section~\ref{incdec}).
Section 4 presents scattering scenarios for the origin of HAT-P-13-like systems,
including a discussion of stellar obliquity in two-planet systems (Section~\ref{twoplanet}).
Section 5 presents a summary,
and Appendix A presents the orbit-averaged equations of motion for 
a two-planet Newtonian point-mass system up to octopole
order, correct to leading order in the inner planet's eccentricity, the ratio of semimajor axes, and the sine
of the inclination of the outer orbit relative to its initial orbit.

\section{Long-term tidal evolution of coplanar systems}\label{coplanar}

In \citet{puffball}, the long-term tidal evolution of short-period planets with single companions is studied. There it is shown
that such systems evolve on three distinct timescales, an illustration of which is
given in Figures 3 and 4 of that paper. 
As long as the eccentricity of the outer planet is non-zero, the eccentricities of both planets will initially execute anti-phased
secular oscillations until a non-zero quasi-equilibrium value is reached, with maxima and minima of the inner planet's
eccentricity
given by expressions (20), (27) or (28) in \citet{puffball}, the choice of which depends on the system parameters,
and the corresponding values of the outer planet's eccentricity given by expression (29).
The equilibrium value of the eccentricity depends on all contributions to the rate of apsidal motion of the 
inner planet. In \citet{puffball} only the contributions from the outer planet and the
post-Newtonian terms in the star's potential were taken into account, however, as \citet{ragozzine} point out,
the contribution of the tidal bulge of a short-period planet like HAT-P-13b is significant and is included here
(as are the contributions from the spin bulge of the planet and the tidal and spin bulges of the star).
The equilibrium eccentricity for a coplanar system is given by
\be
e_b^{(eq)}=\frac{(5/4)(a_b/a_c)\,e_c\,\varepsilon_c^{-2}}
{\left|1-\sqrt{a_b/a_c}(m_b/m_c)\varepsilon_c^{-1}+\gamma\varepsilon_c^3\right|},
\label{equil}
\ee
where $e_b$, $a_b$, and $m_b$ are respectively the 
innermost planet's eccentricity, semimajor axis and mass, and
$e_c$, $a_c$ and $m_c$ are the corresponding values for the outer planet. Here $\varepsilon_c=\sqrt{1-e_c^2}$
and $\gamma=\gamma_*^{GR}+\gamma_b^{tide}+\gamma_b^{spin}+\gamma_*^{tide}+\gamma_*^{spin}$, with
\be
\gamma_*^{GR}=4\left(\frac{n_ba_b}{c}\right)^2\left(\frac{m_*}{m_c}\right)\left(\frac{a_c}{a_b}\right)^3,
\label{gamma1}
\ee
\be
\gamma_b^{tide}=10\, k_b\left(\frac{R_b}{a_b}\right)^5\left(\frac{a_c}{a_b}\right)^3\left(\frac{m_*^2}{m_bm_c}\right),
\hh
\gamma_*^{tide}=10\, k_*\left(\frac{R_*}{a_b}\right)^5\left(\frac{a_c}{a_b}\right)^3\left(\frac{m_b}{m_c}\right),
\label{gamma2}
\ee
\be
\gamma_b^{spin}=\ff{1}{15}\gamma_b^{tide}
\hand
\gamma_*^{spin}=\ff{2}{3}k_*\left(\frac{R_*}{a_b}\right)^5\left(\frac{a_c}{a_b}\right)^3
\left(\frac{m_*}{m_c}\right)\left(\frac{\Omega_*}{n_b}\right)^2
\label{gamma3}
\ee
to first-order in the inner eccentricity.
Here $n_b$ is the mean motion of the inner planet,
$c$ the speed of light, $R_b$ and $k_b$ the radius and Love number of the inner planet,
$m_*$, $R_*$, $k_*$ and $\Omega_*$ the star's mass, radius, Love number and spin frequency respectively,
and synchronous rotation of the planet is assumed in the expression for $\gamma_b^{spin}$.
The $\gamma$'s are the ratios of the various contributions to the apsidal motion of the
inner planet to the coplanar contribution of the outer planet. These are listed in Table~\ref{planet} for the HAT-P-13 system
(for $m_c=m_c^{min}$),
together with those for the star assuming a spin period of 25 days. The Love number for the star (equal to twice its
apsidal motion constant) is taken to
be that for an $n=3$ polytrope \citep{sterne}.

If the angle between the apsidal lines of the two planetary orbits, $\eta$, circulates rather than librates, the average
eccentricity of the inner planet during the initial oscillatory phase will decrease on its own circularization timescale
until its minimum eccentricity is zero, with the amplitude of oscillation remaining constant.
This timescale is given to second order in the eccentricity by
\be
\tau_{circ}=\frac{2}{21n_b}\left(\frac{Q_b}{k_b}\right)\left(\frac{m_b}{m_*}\right)
\left(\frac{a_b}{R_b}\right)^5,
\label{tcirc}
\ee
where $Q_b$ is the $Q$-value of the inner planet.
Once the minimum of the inner eccentricity reaches zero,  $\eta$ will librate about zero or $\pi$
(the choice of which depends on the system parameters), and the oscillation amplitude will reduce to zero until
the inner eccentricity reaches a non-zero quasi-equilibrium value. The librating phase occurs on 
a timescale of $2\tau_{circ}$.

Once the oscillatory phase is over, the system will evolve to the {\it doubly-circular state} on the approximate timescale
\be
\tau_c=\left(\frac{16}{25}\right)\left(\frac{m_c}{m_b}\right)\left(\frac{a_c}{a_b}\right)^{5/2}\cdot
F(e_c^*)\cdot \tau_{circ},
\label{tauc}
\ee
where $e_c^*$ is the value of $e_c$ at the beginning of this phase, and
\be
F(e_c)=\varepsilon_c^{3}(1-\sqrt{a_b/a_c}(m_b/m_c)\varepsilon_c^{-1}
+\gamma\varepsilon_c^3)^{2}\equiv\varepsilon_c^{3}\Delta_0^2.
\ee
Note, however, that if the inner semimajor axis evolves
appreciably on this timescale, \rn{tauc} represents an upper limit only.
Note also that $\tau_c$ is independent of the circularization timescale
of the outer planet; if the latter is comparable to or shorter than
$\tau_c$ then \rn{tauc} again represents an upper bound.
For the HAT-P-13 system, we have $\tau_{circ}=4\times 10^7(Q_b/10^5)$ yr and 
$\tau_c=2.5\times 10^5\tau_{circ}=10,000(Q_b/10^5)$ Gyr, while the orbital decay timescale for planet b \citep{yoder}
is $\tau_a=e_b^{-2}\tau_{circ}\simeq 2500\,\tau_{circ}=100(Q_b/10^5)$ Gyr $<\tau_c$, the latter two being much greater than the age of the system. A direct coplanar integration using the averaged code presented 
in \citet{ML} (using
a constant value of the radius of planet b) gives $\tau_a=79\,(Q_b/10^5)$ Gyr
and $\tau_c=e_c/\dot e_c=9200\,(Q_b/10^5)$ Gyr, while $e_b^{(eq)}$ decays on a timescale $\tau_e=15\,(Q_b/10^5)$ Gyr.\footnote{Note that \citet{ML} code assumes a forcing-frequency-independent $Q$-value,
a legacy of the pioneering work of \citet{goldreich} whose supporting argument was based on
the constancy of the $Q$-value over a wide range of forcing frequencies {\it for the Earth}, ie., for 
a solid body. For gaseous bodies, it seems more reasonable to use the ``constant time-lag'' concept
originally devised by Darwin in which the lag angles of individual tidal components are proportional
to their forcing frequencies. The latter is equivalent to the concept of fluid stress and its associated dissipation.
Note, however, that for small eccentricities (which is the case in the present study),
there is little difference between the formulations.}
The latter is consistent with the estimate 
\be
\tau_e^{-1}=[1+(8\gamma-4\gamma_*^{GR})\varepsilon_c^3/\Delta_0]\tau_a^{-1}\simeq 6\,\tau_a^{-1},
\label{taue}
\ee
obtained from \rn{equil}
with $\Delta_0\simeq1+\gamma\varepsilon_c^3$.
The numerically determined timescales are listed in Table~\ref{planet}, together with the timescale
for decay of the
mutual inclination to its equilibrium value, $\tau_i$, for the case that it is initially $30^\dg$ (see Section~\ref{incdec}).

The estimate for $\tau_e$ places a lower bound on the value of $Q_b$ such that
$Q_b/10^5>({\rm age}/\tau_{e,5})/\ln[e_b(0)/e_b({\rm age})]$, where $\tau_{e,5}=\tau_e(Q_b=10^5)$,
$e_b(0)$ is the ``initial'' value of $e_b$ and $e_b({\rm age})$ is its value now.
Given the lower bound for the age of the system, using $\tau_{e,5}=15$~Gyr and taking $e_b(0)=1$ gives
$Q_b>7000$, while $e_b(0)=0.1$ gives $Q_b>1.8\times 10^4$.
Note that our estimate for $\tau_e$
is more than twice that of \citet{batygin} who estimate $\tau_e\simeq 6 (Q_b/10^5)$~Gyr,
making their lower bound for $Q_b$ a factor 15/6 higher.

In general, coplanar
systems for which one may say something about the structure of the short-period planet will
have the following characteristics:\\
(1) The circularization timescale of planet b will be (considerably) 
less than one third the age of the system, ensuring that the system
is sufficiently relaxed;\\
(2) The timescales on which the system becomes doubly circular and the orbit decays, $\tau_c$ and $\tau_a$
respectively,
will both be longer than the age of the system, and\\
(3) The equilibrium eccentricity and the radius of the short-period planet will be measurable.

\section{Inclined systems}\label{inclined}
In this Section the analysis of \citet{puffball} is generalized to non-zero mutual inclination.
While the HAT-P-13 system is used for illustration, the analysis may be applied to any
system for which $e_b\ll 1$, $a_b/a_c\ll 1$ and 
most of the angular momentum of the system resides in the outer orbit, the latter ensuring that
the (sine of the) angle between the invariable plane and the outer orbit remains small. 

In order to highlight the differences between coplanar and inclined systems,
we begin by presenting the results of an integration of a HAT-P-13-like system for
which the mutual inclination is $30^{\rm o}$, and
the mass of the outer body is taken to be the observed
minimum mass divided by the cosine of the mutual inclination. The stellar obliquity is
such that the star's spin axis is aligned with planet c's orbit normal.
The integration is done using the ``averaged'' code presented
in \citet{ML} which includes accelerations due to spin and tidal bulges of both the star and inner planet
(both quadrupole and dissipative)
as well as the relativistic potential of the star. Orbit-averaged expressions are used for the 
evolution of the inner orbit, while the outer orbit is integrated directly. 
No assumptions are made about the magnitude of any quantity except the ratio of semimajor axes
which is assumed to be small.
For now we suppress the evolution
of the planetary radius and assume it is constant at its currently observed value.

\begin{figure}
\centering
\includegraphics[width=140mm]{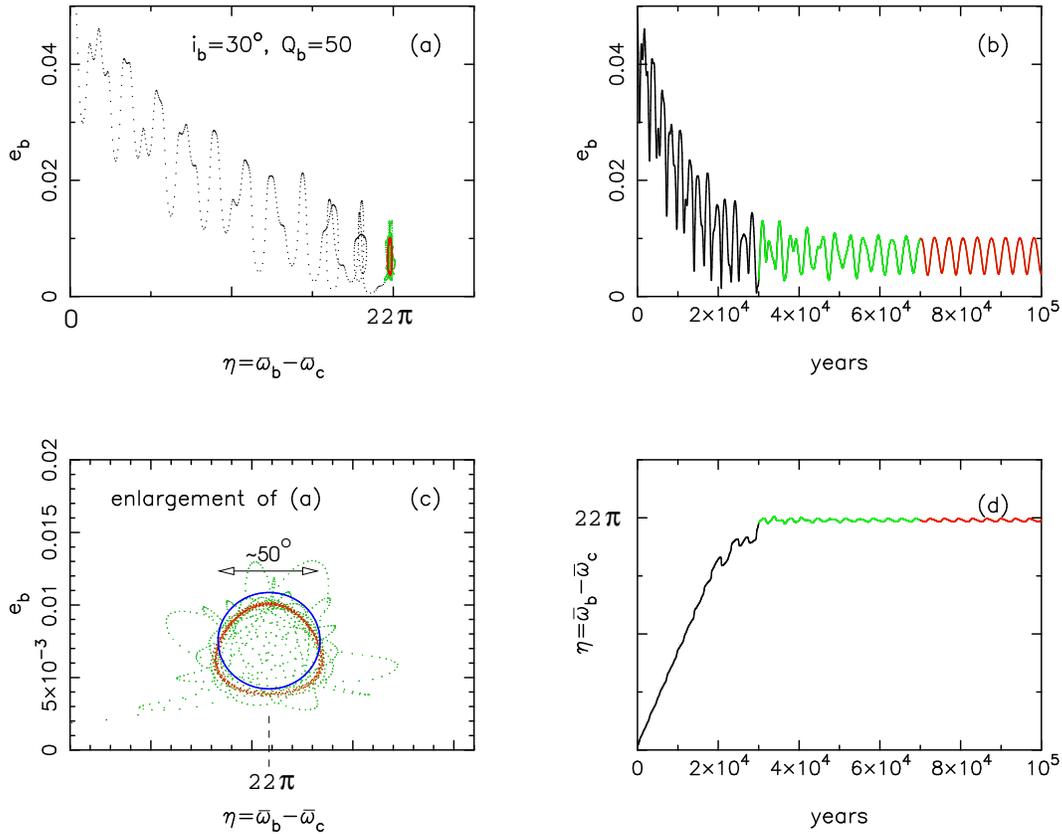}
\caption{The relaxation process for an inclined system (to be compared to Figure~3 in \citet{puffball}).
(a): Capture onto the limit cycle (red points) in $e_b-\eta$ space. Black and green points correspond to
circulation and libration respectively of $\eta$.
(b): Evolution of the eccentricity of planet b. (c): Detail of panel (a); the blue circle corresponds to the analytically
determined limit cycle (Section~\ref{lcprograde}). (d): Evolution of $\eta$.
}
\label{limit-cycle}
\end{figure}

Figure~\ref{limit-cycle}
shows the analogue of Figure~3 in \citet{puffball}. 
The initial values of $\eta=\varpi_b-\varpi_c$ and $e_b$ are $70^\dg$ and 0.05 respectively.
Rather than relaxing to a fixed point in $e_b-\eta$ space on a timescale of $3\tau_{circ}$, the system relaxes
to a limit cycle
on the same timescale. Note that the $Q$-value of planet b is given an
artificially low value of 50 in order to clearly demonstrate capture onto the limit cycle. While the modulation
period is correct, the relaxation process would normally take $Q_b/50$ as long as shown here.
The circulatory phase is shown in black in each panel, the pre-capture libratory phase is shown in green
and the limit-cycle phase is shown in red. Panel (c) shows an enlargement of the limit cycle centred on
$(e_b,\eta)=(e_b^{(av)},22\pi)$, together with the theoretical limit cycle derived below (blue circle).
Here $e_b^{(av)}$ is the inclined-system
analogue of the fixed-point value of $e_b$ for coplanar systems, $e_b^{(eq)}$ (compare equations~\rn{equil}
and \rn{equilinc}). While only around 1.5 relaxation timescales (ie., $4.5\tau_{circ}$)
are shown here, the system was integrated
for 6 relaxation timescales during which $e_b^{(av)}$ and $i_b$ decreased by less than 0.0005
and $0.05^\dg$ respectively.
Note that $20\,\tau_{circ}=2.4 (Q_b/10^5)\,{\rm Gyr}$, comparable to the estimated age of the system
for realistic values of $Q_b$. 

In order to guide the following analysis,
we now present the results of several integrations of relaxed HAT-P-13-like systems with identical
initial conditions except that the mutual inclination is varied between $0^{\rm o}$ and $50^{\rm o}$ (higher inclinations
will be discussed in Sections~\ref{high} and \ref{retrograd}).
The initial value of $e_b$ is taken as $e_b^{(av)}$,
and the apsidal lines are taken to be aligned. 
Initial angles are measured with respect to the
initial orbit of planet c, that is, the relative inclination and longitude of planet b are specified,
with the zero in longitude coinciding with the apse of planet c.
Since planet c's orbit contains 98\% of the total angular momentum, it coincides approximately with the invariable plane
and there is very little change in its inclination as Figure~\ref{all}(e) shows. 
Following \citet{batygin} we take the Love number of planet b to be 0.3, representative of a range
of planetary structures according to their Table~1, and its radius of gyration, $r_{gb}$, to be $0.26R_b$,
appropriate to an $n=1$ polytrope.
For this set of experiments, an artificially low $Q$-value of 10 is used to hasten the evolution towards
the relaxed state. The stellar obliquity {\it relative to the invariable plane normal}, $\theta_*$, is set to zero
initially, and the stellar spin period is 25 days. The stellar Love number and radius of gyration $r_{g*}$
are taken to be 0.03 and $0.076R_*$ respectively, the latter
appropriate to an $n=3$ polytrope.

\begin{figure}
\centering
\includegraphics[width=140mm]{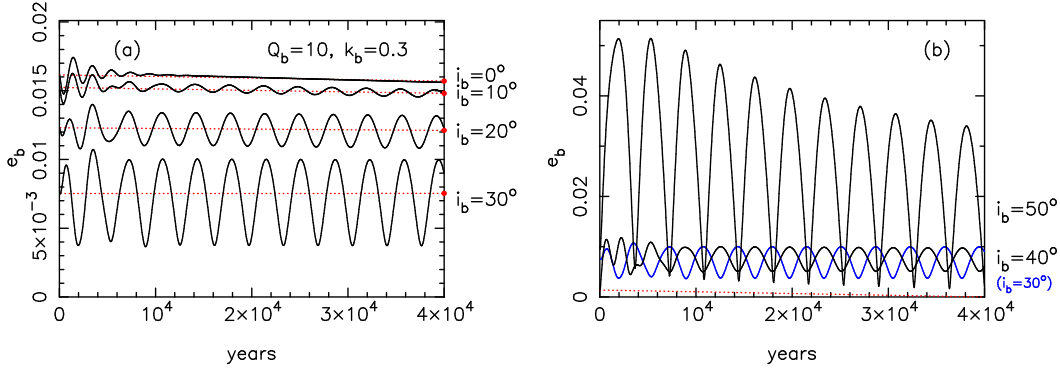}
\caption{The dependence of the relaxed state on
mutual inclination. Note the different scales used for $e_b$ in panels (a) and (b).
(a): The solid curves represent numerically integrated solutions 
for $i_b=0^{\rm o}$, $10^{\rm o}$, $20^{\rm o}$ and $30^{\rm o}$,
while the red dashed
lines are the instantaneous theoretical average values of $e_b$,
$e_b^{(av)}$, given by equation~\rn{equilinc}.
The artificially low $Q$-value allows the transient behaviour to die away quickly. Note
that the modulation frequency and amplitude are independent of $Q_b$ and are given by $2W_\omega$
and $A/2W_\omega$,
with $W_\omega$ and $A$ defined in equations~\rn{Ww} and \rn{A} respectively.
(b): Integrated solutions for $i_b=40^{\rm o}$ and $50^{\rm o}$ (black curves) together with $i_b=30^{\rm o}$
for comparison (blue curve). The cases $i_b=30^{\rm o}$ and $40^{\rm o}$ have similar averages,
while the amplitude of variation of the $i_b=50^{\rm o}$ system is relatively large. Its decay
is associated with the relatively rapid decay of the semimajor axis of planet b for that case.
Also shown is $e_b^{(av)}$ for $i_b=40^{\rm o}$ (red dotted line); clearly equation~\rn{equilinc}
is not reliable for this case.
}
\label{eeq4}
\end{figure}

\begin{figure}
\centering
\includegraphics[width=160mm]{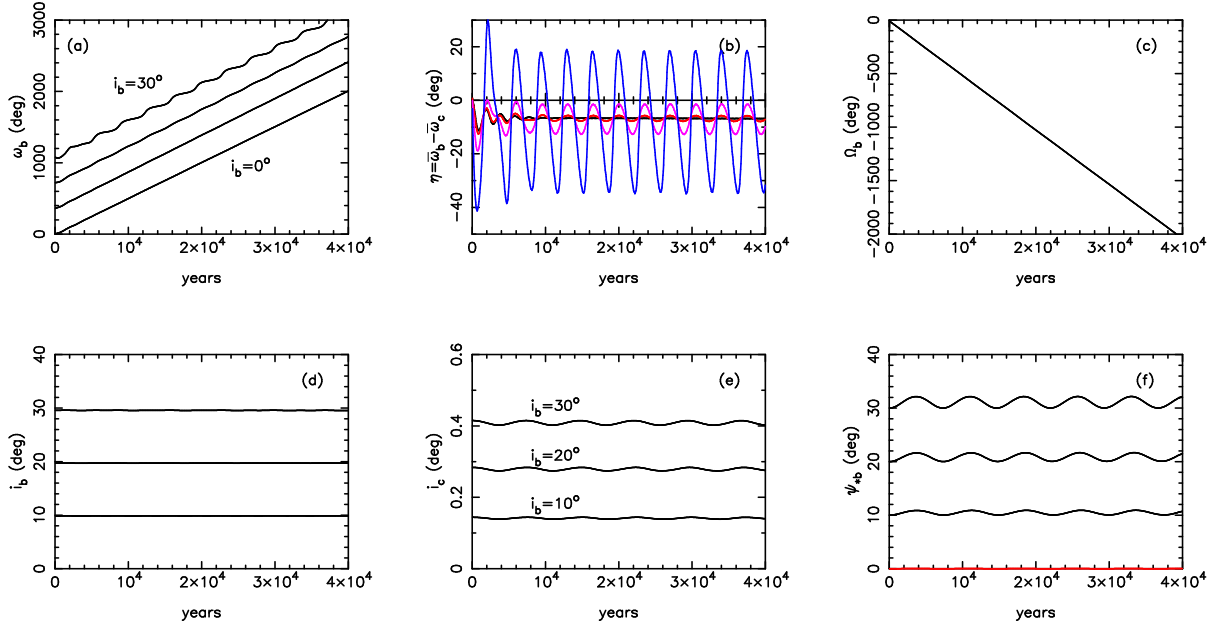}
\caption{Variation of the elements for $i_b=0^{\rm o}$, $10^{\rm o}$, $20^{\rm o}$ and $30^{\rm o}$.
Colours in panel (b) are such that black, red, pink and blue correspond to $0^\dg$, $10^\dg$, $20^\dg$ and $30^\dg$
respectively.
See text for a discussion of each panel.}
\label{all}
\end{figure}

Figure~\ref{eeq4}(a)
compares the evolution of $e_b$ for $i_b=0^{\rm o}$, $10^{\rm o}$, $20^{\rm o}$ and $30^{\rm o}$. The two main
features of this plot are that the average eccentricity decreases with increasing mutual inclination,
while the amplitude of its oscillation increases. Both these quantities can be determined from the
analysis below, as can the modulation period of $e_b$. Note that all three are independent of the $Q$-value 
of the planet (for small $\sin\eta$ and constant planet radius) so that Figure~\ref{eeq4} represents their true values.
Figure~\ref{eeq4}(b) shows the evolution of $e_b$ for $i_b=40^{\rm o}$ and $50^{\rm o}$ (black curves)
as well as for $i_b=30^{\rm o}$ for comparison (blue curve). 

Figure~\ref{all}
shows the evolution of the other orbital elements for each value of $i_b$.
Panel (a) shows the evolution of the {\it argument} of periastron, each offset by
$360^{\rm o}$ for clarity. Note in particular that for these initial inclinations, $\dot\omega_b$
is approximately constant, and is given by \rn{Ww}. 
Panel (b) shows the libratory behaviour of $\eta$.
Unlike the coplanar case which evolves to a constant value such that $\sin\eta=-W_T/|W_q|$
(and hence is proportional to $Q_b^{-1}$), the quasi-relaxed state for systems with non-zero mutual inclination
is oscillatory (although the offset is still ${\rm Sin}^{-1}[-W_T/|W_q|])$.
The amplitude of oscillation is given by
$A/2W_\omega$ (times $180/\pi$),
with $A$ and  $W_\omega$ defined in equations~\rn{A} and \rn{Ww} respectively.

Panel (c) shows the evolution of the longitude of the ascending node of planet b for {\it all} inclinations.
Note that while planet b's precession rate is proportional to $m_c\cos i_b$ (equation~\rn{Ob}), $m_c=m_c^{min}/|\cos i_b|$
with $m_c^{min}$ held constant so that the precession rates are all identical. The precession period is 7281 years.
Panels (d) and (e) show the evolution of the inclinations of orbits b and c relative to the invariable plane, $i_b$,
and $i_c$ respectively,
justifying our assumption in the analysis below that they are constant and that $\sin i_c\ll 1$
(see Section~\ref{twoplanet} for analysis of the influence on $i_b$ of non-zero $\theta_*$).
Not shown is the variation in planet c's eccentricity; this is effectively constant with an amplitude of
variation of 0.003.

Finally, panel (f) shows the stellar obliquity relative to orbit b (the angle between the star's spin axis and planet b's orbit normal),
$\psi_{*b}$.
In Section~\ref{twoplanet} we show that the maxima and minima of $\psi_{*b}$ are $|i_b\pm\theta_*|$.
In fact, $\theta_*$ varies (the star nutates) due to torques between the misaligned stellar spin bulge and b's orbit,
consistent with the variations seen in panel (f).

In light of the recent discoveries of high (sky-projected) stellar obliquities in
the systems HAT-P-7 \citep{narita,winn} and WASP-17 \citep{anderson}, we discuss further the dynamics
of stellar obliquity in Section~\ref{twoplanet}.

\subsection{Limit-cycle behaviour of prograde orbits}\label{lcprograde}

In Appendix A we give the orbit-averaged disturbing function for a two-planet Newtonian point-mass system up to octopole
order for arbitrary planet b inclination, and from this derive the equations governing the secular evolution
of the elements in the absence of perturbations. 
Only leading order terms in $e_b$, $a_b/a_c$ and $\sin i_c$ are retained, each of which are
of order 0.01 for the HAT-P-13
system whether or not the reference plane is taken as the initial plane of planet c or the invariable plane.
Equations \rn{ep}, \rn{wp}, \rn{ec} and \rn{wc} should be compared with equations (4)-(7) of \citet{puffball} for the coplanar case,
noting that here the inclination functions are such that $f_n(0)=1$, $n=1,2,3$ and $g_2(0)=0$.

Our aim now is to write down equations governing the dominant long-term 
behaviour of the two-planet system under the action
of tidal dissipation, spin-orbit coupling and the relativistic potential of the star.
In particular, we wish to study the behaviour of the system in the $e_b-\eta$ plane in order to 
determine whether a relaxed system with non-zero mutual inclination
is able to tell us anything about the internal structure of planet b.
Noting from Figure~\ref{all} (and the following analysis) that
$\dot\omega_b=\dot\varpi_b-\dot\Omega_b$ is approximately constant, and that
as long as $\eta$ librates, the argument $\zeta\equiv\varpi_b+\varpi_c-2\Omega_b=2\omega_b-\eta$ so that
$\dot\zeta\simeq 2\dot\omega_b$,
the equations governing $e_b$ and $\eta$ are approximately
\be
\dot e_b=-\left[W_T\, e_b+W_o \,e_c\,\sin\eta\right]+A\,e_b\sin (2\dot\omega_b t)+B\,e_c\sin(2\dot\omega_b t),
\label{eb2}
\ee
and
\be
\dot\eta=\left[W_q-W_o\left(\frac{e_c}{e_b}\right)\cos\eta\right]+A\,\cos (2\dot\omega_b t)+
B\left(\frac{e_c}{e_b}\right)\cos(2\dot\omega_b t),
\label{eta}
\ee
with
\be
\dot\omega_b=\dot\eta+\dot\varpi_c-\dot\Omega_b\simeq \dot\varpi_c-\dot\Omega_b,
\label{omega2}
\ee
and we have taken the time origin to coincide with $\omega_b=0$.
Here
$W_T=\tau_{circ}^{-1}$ is the inverse of the tidal circularization timescale,
and both $W_q$ and $W_o$ have analogues in the coplanar theory\footnote{The subscripts
$q$ and $o$ stand for quadrupole and octopole respectively.} and are such that
\be
W_q=
\left[f_3(i_b)-\left(\frac{m_b}{m_c}\right)\sqrt{\frac{a_b}{a_c}}\varepsilon_c^{-1}\cdot f_1(i_b)+\gamma\varepsilon_c^3\right]\,
W_\Omega\equiv \Delta \cdot W_\Omega
\label{Wq}
\ee
and
\be
W_o=\frac{5}{4}\left(\frac{a_b}{a_c}\right)\varepsilon_c^{-2}\, f_2(i_b)\,W_\Omega,
\ee
with
\be
W_\Omega=\frac{3}{4}n_b\left(\frac{m_c}{m_*}\right)\left(\frac{a_b}{a_c}\right)^3 \varepsilon_c^{-3}
\label{WOmega}
\ee
equal to minus the precession frequency divided by $\cos i_b$
(see equation~\rn{Ob}). 
Parameters which are zero for coplanar systems are
\be
A=\ff{5}{2}\, \sin^2 i_b\,W_\Omega
\hand
B=\frac{5}{4}\left(\frac{a_b}{a_c}\right)\varepsilon_c^{-2}\, g_2(i_b)\,W_\Omega,
\label{A}
\ee
while
\be
\dot\omega_b\simeq
\left[\cos i_b+\left(\frac{m_b}{m_c}\right)\sqrt{\frac{a_b}{a_c}}\varepsilon_c^{-1}\cdot f_1(i_b)\right]\,W_\Omega
\label{Ww}
\ee
does not appear explicitly in the coplanar theory.
Note that since $a_b/a_c$, $i_b$, $i_c$ and $e_c$ are all approximately constant over a few times
the circularization timescale (at least for small $e_b$), each of $W_T$, $W_q$, $W_o$, $W_\Omega$,
$\dot\omega_b$, $A$ and $B$ are also approximately constant on this timescale.

\begin{figure}
\centering
\includegraphics[width=170mm]{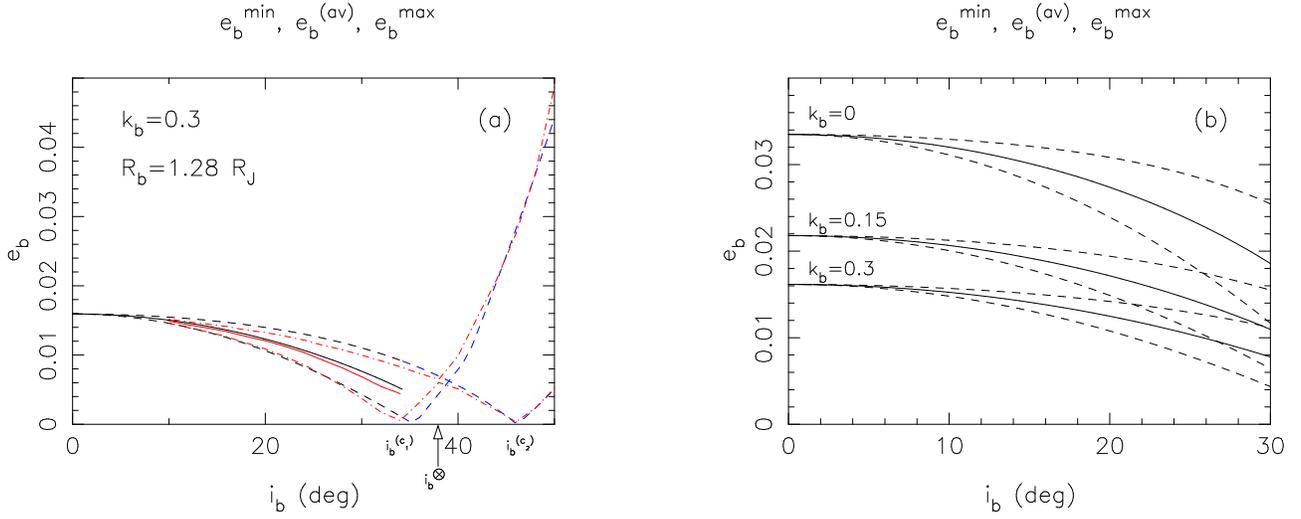}
\caption{(a): Comparison of theoretical (black curves) and numerical values (red curves)
of $e_b^{(av)}$ (solid curves), $e_b^{min}$ and $e_b^{max}$ (dashed curves for theoretical and dot-dashed curves
for numerical)
as functions of $i_b$. The mass of planet c is $m_c^{min}/\cos i_b$.
Note that we have scaled the expression for ${\cal A}_{lc}$ given by
\rn{Alc} by a factor of two thirds
to improve the fit.
The theoretical estimates for $e_b^{min}$ and $e_b^{min}$ are listed in Table~\ref{ebminmax},
while that for $e_b^{(av)}$ is given by \rn{equilinc}. Note that $\eta$ librates around zero for $0\leq i_b<i_b^{(c_1)}$,
circulates for $i_b^{(c_1)}<i_b<i_b^{(c_2)}$
and librates about $\pi$ for $i_b>i_b^{(c_2)}$. 
(b): Theoretical estimates for the dependence of $e_b^{(av)}(i_b)$, $e_b^{min}(i_b)$ and $e_b^{max}(i_b)$ on $k_b$.
Since the amplitude of variation of $e_b$ in the relaxed state is small for $i_b\lapp 10^{\rm o}$,
a measurement of $e_b$ gives a fairly accurate estimate of $k_b$ for those inclinations.
}
\label{eeq-comp-k}
\end{figure}

Referring to Figure~\ref{eeq4} showing the relaxed behaviour of $e_b$ for different $i_b$,
our aim now is to characterize the limit cycle demonstrated in Figure~\ref{limit-cycle} by
determining the average value of $e_b$ as well as its amplitude as a function of the
mutual inclination. Equations~\rn{eb2} and \rn{eta} together represent a damped nonlinear non-homogeneous 
system of ordinary differential equations with periodic coefficients for which the existence of periodic (limit-cycle) solutions are
suggested by Figures~\ref{eeq4} and \ref{all} (see, for example, \citet{jordan} for a discussion of such systems). 
Assuming a limit-cycle frequency of $2\dot\omega_b$ and zero phase,\footnote{
One can alternatively leave the frequency and phase as parameters to be determined using the procedure described here.} 
that $\eta$ librates rather than circulates (this is true for $i_b\lapp 33^{\rm o}$ for the HAT-P-13 system
as determined from numerical integrations),
and that relaxed values for $e_b$ and $\eta$ are $\pi/2$ out of phase (a reasonable assumption given the form of 
\rn{eb2} and \rn{eta}),
the amplitude of variation of $e_b$ and $\eta$ on the limit cycle can be estimated by putting
\be
e_b^{(lc)}(t)=e_b^{(av)}\left[1-{\cal A}_{lc}\cos(2\dot\omega_b t)\right]
\hand
\eta^{(lc)}(t)=\eta_{av}+{\cal A}_{lc}\sin(2\dot\omega_b t),
\label{eA}
\ee
where $e_b^{(av)}$ and $\eta_{av}$ are defined in the following procedure. Substituting these expressions into 
\rn{eb2} and \rn{eta}, assuming ${\cal A}_{lc}\ll 1$ and $\eta^{(lc)}\ll 1$,
matching constant terms and those with phase $2\dot\omega t$ and neglecting terms with phase
$4\dot\omega t$, we obtain 
\be
{\cal A}_{lc}=\frac{A+B e_c/e_b^{(av)}}{2\dot\omega_b+W_o e_c/e_b^{(av)}}
=\frac{A+(g_2/f_2)W_q}{2\dot\omega_b+W_q}
=
\frac{\ff{5}{2}\sin^2 i_b+(g_2/f_2)\Delta}
{2[\cos i_b+\sqrt{a_b/a_c}(m_b/m_c) \varepsilon_c^{-1}f_1(i_b)]+\Delta},
\label{Alc}
\ee
where $\Delta$ is defined in \rn{Wq}, together with
\be
e_b^{(av)}=(W_o/W_q)e_c=\frac{(5/4)(a_b/a_c)\,e_c\,\varepsilon_c^{-2}\cdot f_2(i_b)}
{f_3(i_b)-\sqrt{a_b/a_c}(m_b/m_c)\varepsilon_c^{-1}\cdot f_1(i_b)+\gamma\varepsilon_c^3}
\label{equilinc}
\ee
which reduces to \rn{equil} when $i_b=0$, and 
\be
\eta_{av}=-W_T/W_q=-\frac{7}{5}\frac{\gamma_b^{tide}}{Q_b}\frac{\varepsilon_c^3}{\Delta}.
\label{eav}
\ee
Expressions \rn{equilinc} and \rn{eav} should be compared with equations (36) and (46) in \citet{puffball}.
The minimum and maximum values of $e_b^{(lc)}$ are therefore
\be
e_b^{min,max}=e_b^{(av)}\left(1\pm {\cal A}_{lc}\right).
\label{minmax}
\ee
In fact the true amplitude tends to be around $\ff{2}{3}{\cal A}_{lc}$, perhaps due to the neglect
of terms proportional to $\cos (4\omega_b)$ and $\sin (4\omega_b)$ (these are associated with the shape of the 
true limit cycle - see Figure~\ref{limit-cycle}). We therefore replace ${\cal A}_{lc}$ with
\be
{\cal A}_{lc}^*\equiv\ff{2}{3}{\cal A}_{lc}
\ee
in the analysis and figures that follow (as well as Figure~\ref{limit-cycle}).
Figure~\ref{eeq-comp-k}(a) 
plots $e_b^{(av)}$ from \rn{equilinc} (black solid curve) together with $e_b^{min}$ and $e_b^{max}$ 
from \rn{minmax} (black dashed curves)
for $i_b\lapp 33^\dg\equiv i_b^{(c_1)}$, the range of values of $i_b$ corresponding to libration of $\eta$ around $\eta=\eta_{av}$.
Also plotted are maximum, minimum and average values of $e_b$ from
a series of numerical integrations for HAT-P-13-like systems with $10^{\rm o}\leq i_b\leq 50^{\rm o}$
(red solid and dot-dashed curves). 
The value $i_b=i_b^{(c_1)}$ corresponds to ${\cal A}_{lc}^*=1$ (compare with the discussion following
equation~(23) in \citet{puffball}). For $i_b^{(c_1)}\lapp i_b\lapp i_b^{(c_2)}\simeq 46^\dg$, $\eta$ {\it circulates},
these values of $i_b$ corresponding to $|{\cal A}_{lc}^*|>1$, while for $i_b^{(c_2)}<i_b<i_b^K\simeq 54^\dg$,
$\eta$ again librates, this time around $\eta=\pi+\eta_{av}$. Here $i_b^K$ corresponds to the minimum value of $i_b$
for which Kozai oscillations occur for a given value of $\gamma$ (see Section~\ref{high}).

While it is straightforward to determine $e_b^{min}$ and $e_b^{max}$ for all possible (non-relaxed) coplanar configurations 
and their associated fixed points \citep{puffball},
it not clear how to do this for inclined systems and their associated limit cycles for systems with $i_b>i_b^{(c_1)}$,
that is, systems which $\eta$ either circulates or librates around $\pi$.
Here we content ourselves with trial and error expressions, obtained by trying different integer values
of $n_1$ and $n_2$ in $e_b^{(av)}(n_1{\cal A}_{lc}+n_2)$, and comparing these with numerical solutions.
The results are summarized in Table~\ref{ebminmax}
\begin{table*}
 \begin{minipage}{170mm}
  \caption{Eccentricity maxima and minima.}
  \label{ebminmax}
  \begin{tabular}{lccccr}
  \hline
& $e_b^{min}$ & $e_b^{max}$ & & & behaviour\\ \hline
$0\leq i_b< i_b^{(c_1)}$ \phantom{aa} & \phantom{aa}$e_b^{(av)}(1-{\cal A}_{lc}^*)$  \phantom{aa} 
&  \phantom{aa} $e_b^{(av)}(1+{\cal A}_{lc}^*$)  
 \phantom{aa}  & \phantom{aa} $e_b^{(av)}>0$  \phantom{aa} & $0\leq{\cal A}_{lc}^*<1$ & libration around 0\\
$i_b^{(c_1)}\leq i_b< i_b^\otimes$ & $2\,e_b^{(av)}({\cal A}_{lc}^*-1)$ & $e_b^{(av)}({\cal A}_{lc}^*+1)$ & $e_b^{(av)}>0$ 
& ${\cal A}_{lc}^*\geq1$ & circulation\\
$i_b^\otimes\leq i_b< i_b^{(c_2)}$ & $e_b^{(av)}({\cal A}_{lc}^*+1)$ & $2\,e_b^{(av)}({\cal A}_{lc}^*-2)$ 
& $e_b^{(av)}<0$ & ${\cal A}_{lc}^*<-1$ & circulation\\
$i_b^{(c_2)}\leq i_b< i_b^{K} $ & $-e_b^{(av)}({\cal A}_{lc}^*+1)$ &  $2\,e_b^{(av)}({\cal A}_{lc}^*-2)$ & $e_b^{(av)}<0$ & $-1\leq{\cal A}_{lc}^*<0$ &  \phantom{aa}  libration around $\pi$\\
\hline
\end{tabular}
\end{minipage}
\end{table*}
 and plotted in Figure~\ref{eeq-comp-k}(a), the latter (blue dashed curves) showing good agreement with numerical solutions (red dashed curves).
The angle $i_b^\otimes\simeq 39^\dg$ corresponds to the point where $e_b^{(av)}$ and
${\cal A}_{lc}^*$ change sign and as a consequence, the $e_b^{min}$ and $e_b^{max}$ curves cross.
Given the form of these expressions and the accuracy with which they fit the numerical data, it seems likely that they are generic.

Figure~\ref{eeq-comp-k}(b) demonstrates the dependence of 
$e_b^{(av)}$, $e_b^{min}$ and $e_b^{max}$ on the Love number of planet b. Note that for the case
$k_b=0$ we have also set $k_*$ to zero (although in the $k_b=0.3$ case the quadrupole moment of the
star contributes only 1.4\% to the total value of $\gamma$, assuming the star spins with the same period as the Sun).
{\it We conclude that 
the analysis and conclusions of \citet{batygin} are valid as long as the mutual inclination of planets b and c
is less than around $10^\dg$; for higher values of $i_b$, a measurement of $e_b$ does not unambiguously
determine $k_b$}, although one can make arguments about the likelyhood of a system being near
the top or bottom of the modulation cycle of $e_b$. While a  Rossiter-McLaughlin estimate of the stellar obliquity
will not help to constrain $i_b$ (see discussion in Section~\ref{twoplanet}), it should be possible to 
fit combined transit and spectroscopic data for the mutual inclination \citep{nesvorny}.

We now argue that for realistic values of $Q_b$, the HAT-P-13 system cannot have a mutual inclination  
in the range $54^\dg$ to $126^\dg$.

\subsection{The Kozai regime}\label{high}

\begin{figure}
\centering
\includegraphics[width=150mm]{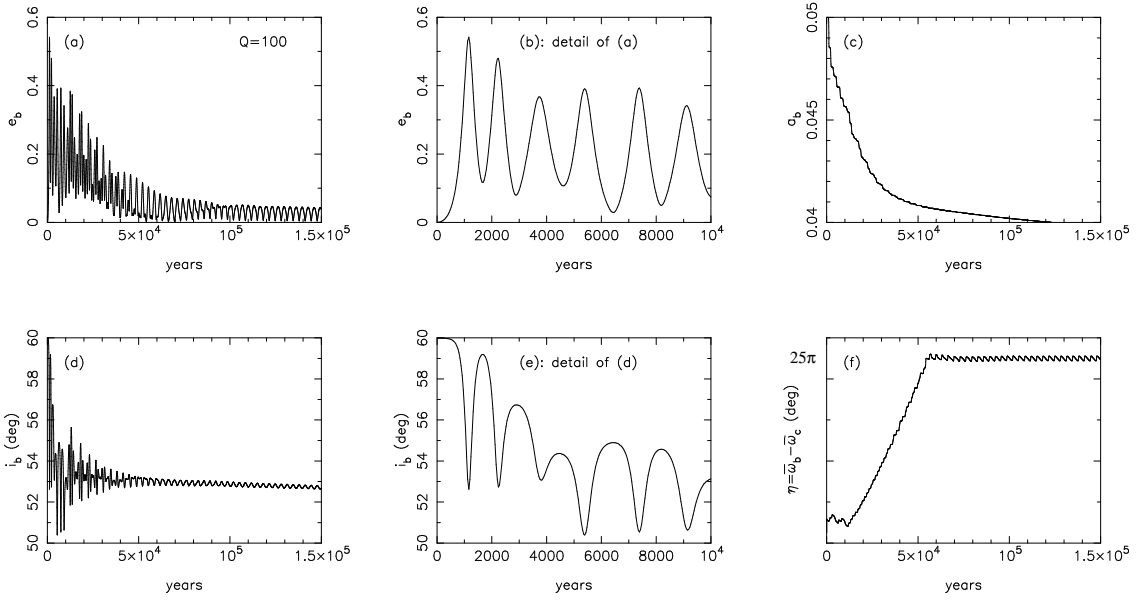}
\caption{The Kozai regime. Here the initial mutual inclination and eccentricity of planet b are $60^\dg$ and zero
respectively, and $Q_b=100$. The eccentricity of planet b is forced to high values initially (panel (b))
and there are corresponding variations in $i_b$ (panel (e)),
however, this extreme behaviour comes at the price of severe tidal decay of the orbit of planet b (panel (c))
and the mutual inclination (panel (d)) until the system is no longer capable of driving Kozai oscillations. This occurs
at around $i_b=53^\dg$ at which point the system becomes trapped on a limit cycle for which $\eta=25\pi+\eta_{av}$
and $e_b^{av}\simeq 0.02$. This value of $i_b=i_b^K$ (see Table~\ref{ebminmax}) 
is consistent with the analysis summarized in Figure~\ref{kozai}.
Note that $m_c=m_c^{min}/\cos 60^\dg$.
}
\label{allk}
\end{figure}

Figure~\ref{allk}
shows the evolution of $e_b$, $a_b$, $i_b$ and $\eta=\varpi_b-\varpi_c$ for the
case $i_b(t=0)=60^{\rm o}$, $e_b(0)=0$, $a_b(0)=0.05$ and $Q_b=100$ (all other parameters are the same as for the
cases presented in Section~\ref{inclined}). 
The system exhibits Kozai oscillations, detail of which is shown in panels (b) and (e) for $e_b$ and $i_b$
respectively. However, given that there is no commensurate variation of $a_b$,
these come at the price of severe tidal decay of the orbit of planet b as shown
in panel (c). An upper bound for the timescale for decay of the semimajor axis of planet b's orbit due to planetary tides is
$\tau_a=\tau_{circ}/e_b^2$,
where $\tau_{circ}$ is given by \rn{tcirc}. With $e_b$ at least an order of magnitude higher than for the
cases considered in Section~\ref{inclined}, here orbital decay occurs more than one hundred times faster during the
high-eccentricity phase for a given value of $Q_b$
(note that for this example we have used a $Q$-value 10 times as large to clearly demonstrate Kozai cycles).
The effect of this strong tidal damping is to quickly reduce the mutual inclination
to a value for which Kozai oscillations no longer occur (around $53^{\rm o}$; see discussion below) and the apsidal lines become
locked with $\eta\simeq 25\pi$ (panel (f)). The start of the slower phase corresponds to
a value of planet b's semimajor axis of around 0.041 AU (the observed value is 0.0426 AU).
Thereafter there is a slow decline in $i_b$, $a_b$ and $e_b$. 

\begin{figure}
\centering
\includegraphics[width=170mm]{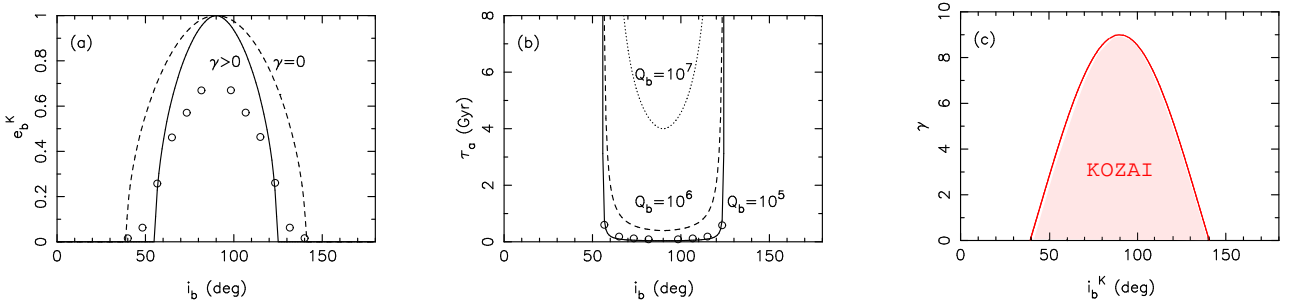}
\caption{(a): Maximum (Kozai) eccentricity when only quadrupole terms are included as a function of inclination
and $\gamma$. 
The dashed black curve corresponds to $\gamma=0$, while
the solid black curve, appropriate to the HAT-P-13 system with $k_b=0.3$, 
corresponds to systems with $m_c=m_c^{min}/|\cos i_b|$ 
so that $\gamma(m_c=m_c^{min}/|\cos i_b|)=\gamma(m_c=m_c^{min})\cdot|\cos i_b|$ 
(see equations~\rn{gamma1}-\rn{gamma3})
decreases along both sides of the curve to zero at $i_b=90^\dg$. The maximum value of $\gamma$
is 4.2 at the end-points $i_b=54^\dg$ and $126^\dg$.
The circles correspond to the ``true'' maximum values of $e_b$ for the $k_b=0.3$ case, 
calculated using the averaged code so that all tidal and octopole terms are included.
(b): Orbit decay timescale as a function of $i_b$. The solid curve and circles
correspond to the solid curve and circles in panel (a) with $Q_b=10^5$, while the dashed 
and dotted curves
corresponds to the same but with $Q_b=10^6$ and $Q_b=10^7$ respectively. We conclude from this that the HAT-P-13 system
is unlikely to have a mutual inclination between $54^\dg$ and $126^\dg$ for reasonable values of $Q_b$.
(c): The range of inclinations for which Kozai oscillations of $e_b$ occur
given a particular value of $\gamma$. 
}
\label{kozai}
\end{figure}

\citet{fabrycky} give an expression for the maximum eccentricity, $e_b^{K}$, attained during 
a Kozai cycle in the presence of other perturbing forces such as spin, tides and relativity.
Their equation (34) (which holds in the case that the initial inner eccentricity is zero)
can be rearranged so that 
\be
e_b^{K}=\sqrt{1-x^2},
\label{FT}
\ee
where $x$ is the minimum positive root of the cubic
\be
x^3+x^2-(C_1+C_2)x-C_1=0,
\ee
with
$C_1=\ff{5}{3}\cos^2 i_b(0)$ and $C_2=\ff{2}{9}\gamma$, $i_b(0)$ being the initial inclination
of planet b (which equals the initial relative inclination since $i_c(0)=0$).
Note that this expression does not include any contribution from octopole terms in the disturbing function,
nor does it include terms of order $e_b^2$ or higher
in the tidal distortion.
Figure~\ref{kozai}(a)
shows the dependence of $e_b^K$ on $i_b(0)$ (solid curve), taking into account the fact that
the minimum mass of planet c is scaled by $(\cos i_b)^{-1}$ so that the apsidal advance of planet b is
completely dominated by planet c when $i_c=90^\dg$ (see also Figure~3 of \citet{fabrycky}).
The circles correspond to ``true'' maximum values of $e_b$ for this case, 
calculated using the averaged code (with no dissipation) so that all tidal and octopole terms are included.
While the maximum and minimum values of $e_b$ do not vary from one Kozai cycle to the next when only 
point-mass quadrupole terms are included in the equations of motion, this is not true when extra
accelerations are included, although the variation is generally small (less than 10\%). 
We therefore integrated for several modulation cycles and recorded the maximum value of $e_b$ during
that time.
The results suggest that the terms not included in \rn{FT} make a significant difference for high inclinations.
Also shown in panel (a) is $e_b^{K}$ when $\gamma=0$ (dashed curve).
Figure~\ref{kozai}(b) shows the dependence of the orbit decay timescale on mutual inclination, for which an estimate
is given by
\be
\tau_a=(e_b^K)^{-2}\tau_{circ}.
\label{tauaK}
\ee
That this gives a good estimate (rather than using, say, the average value of $e_b$ during a Kozai cycle)
is supported by the example in Figure~\ref{allk} for which panel (c) provides an estimate over the first
$10^4$~yr of $\tau_a=1.25\times 10^5$~yr while \rn{tauaK} with $e_b^K=0.4$ (panel (b)) gives $\tau_a=2.5\times 10^5$~yr,
that is, \rn{tauaK} represents an upper bound.
The circles in panel (b) of Figure~\ref{kozai} 
correspond to those in panel (a) and show that \rn{FT} may be used in \rn{tauaK} to estimate
$\tau_a$ in spite of the fact that not all relevant terms are included. We may therefore conclude from panel (b) that
{\it given an estimated age of around 5 Gyr,
the mutual inclination of the HAT-P-13 system cannot be between 54 and 126 deg
for values of $Q_b$ less than $10^6$}. Note that if $Q_b$ is as high as $10^7$, the system will not yet
have relaxed to the limit-cycle state.

Finally, in order to be able to place bounds on the mutual inclination of any given observed system, 
it is important to know the range of inclinations for which Kozai oscillations of $e_b$ occur
given a particular value of $\gamma$.
Figure~\ref{kozai}(c)
shows that Kozai oscillations are completely suppressed for $\gamma\gapp 9$ for
any inclinations. This also has repercussions for the maximum possible stellar obliquity of a system, discussed in 
Section~\ref{kozai}.

\subsection{Limit-cycle behaviour of retrograde systems}\label{retrograd}

While relaxed prograde systems are characterized by the libration of the angle $\eta=\varpi_b-\varpi_c$,
relaxed retrograde systems are characterized by libration of the angle 
$\zeta=\varpi_b+\varpi_c-2\Omega_b$. This can be understood as follows.

Noting that $\eta=2\omega_b-\zeta$, and that $\dot\eta\simeq 2\dot\omega_b$ when $\zeta$ librates,
equations~\rn{eb2}-\rn{omega2} can be written
\be
\dot e_b=-\left[W_T\, e_b+W_o^r \,e_c\,\sin\zeta\right]+A\,e_b\sin (2\dot\omega_b t)+B^r\,e_c\sin(2\dot\omega_b t),
\label{eb2r}
\ee
and
\be
\dot\zeta=\left[W_q^r-W_o^r\left(\frac{e_c}{e_b}\right)\cos\zeta\right]+A\,\cos (2\dot\omega_b t)+
B^r\left(\frac{e_c}{e_b}\right)\cos(2\dot\omega_b t),
\label{zeta}
\ee
with
\be
\dot\omega_b=\dot\zeta-\dot\varpi_c+\dot\Omega_b\simeq -\dot\varpi_c+\dot\Omega_b,
\label{omega2r}
\ee
where
\be
W_q^r=
\left[h_3(i_b)-\left(\frac{m_b}{m_c}\right)\sqrt{\frac{a_b}{a_c}}\varepsilon_c^{-1}\cdot f_1(i_b)+\gamma\varepsilon_c^3\right]\,
W_\Omega
\label{Wqr}
\ee
and
\be
W_o^r=\frac{5}{4}\left(\frac{a_b}{a_c}\right)\varepsilon_c^{-2}\, g_2(i_b)\,W_\Omega,
\ee
with $W_\Omega$ defined in \rn{WOmega}.
Parameters which are zero for retrograde coplanar systems are
\be
A=\ff{5}{2}\, \sin^2 i_b\,W_\Omega
\hand
B^r=\frac{5}{4}\left(\frac{a_b}{a_c}\right)\varepsilon_c^{-2}\, f_2(i_b)\,W_\Omega,
\label{Ar}
\ee
while
\be
\dot\omega_b\simeq
-\left[\cos i_b+\left(\frac{m_b}{m_c}\right)\sqrt{\frac{a_b}{a_c}}\varepsilon_c^{-1}\cdot f_1(i_b)\right]\,W_\Omega.
\label{Wwr}
\ee
Note that the rate of 
precession of the node of planet b about
the invariable plane normal is negative for prograde orbits and positive for retrograde orbits (equation~\rn{Ob}).
Given the symmetry properties of the various inclination functions (listed in the paragraph
following \rn{A10}), we see that the relaxation behaviour of those retrograde systems for which $\zeta$ librates
is identical to that of prograde systems for which $\eta$ librates (albeit with a slightly different limit cycle frequency).
In particular, the theory developed in Sections~\ref{lcprograde} and \ref{incdec}
for prograde systems caries directly over to retrograde systems if one replaces
$f_2(i_b)$, $f_3(i_b)$ and $f_4(i_b)$ with
$g_2(i_b)$, $h_3(i_b)$ and $h_4(i_b)$ respectively, as well as $\dot\omega_b$ with $-\dot\omega_b$.
In particular, $e_b^{(av)}(i_b)=e_b^{(av)}(\pi-i_b)$ and ${\cal A}_{lc}(i_b)={\cal A}_{lc}(\pi-i_b)$, while
$\langle di_b(i_b)/dt\rangle=-\langle di_b(\pi-i_b)/dt\rangle$ (due to the factor $\sin(2i_b)$).
Note, however, that $\dot\omega_b(i_b)\neq-\dot\omega_b(\pi-i_b)$ because while $f_1(i_b)=f_1(\pi-i_b)$,
$\cos i_b=-\cos(\pi-i_b)$.

\subsection{The relationship between $e_b^{(av)}$, the equilibrium radius of 
planet b, its $Q$-value and its core mass}\label{seceReq}

The above analysis implicitly assumes that the value taken for $R_b$ is its equilibrium value.
In fact, for a given set of orbital parameters and planetary core mass, the equilibrium value of $R_b$,
$R_b^{(eq)}$, depends on $Q_b$. This, in turn, determines $e_b^{(av)}$ for a given value of $i_b$.
Moreover, the Love number depends on the core mass as well as the radius.
We can quantify these relationships using the data provided in Table 1 of \citet{batygin} as follows.

\begin{figure}
\centering
\includegraphics[width=140mm]{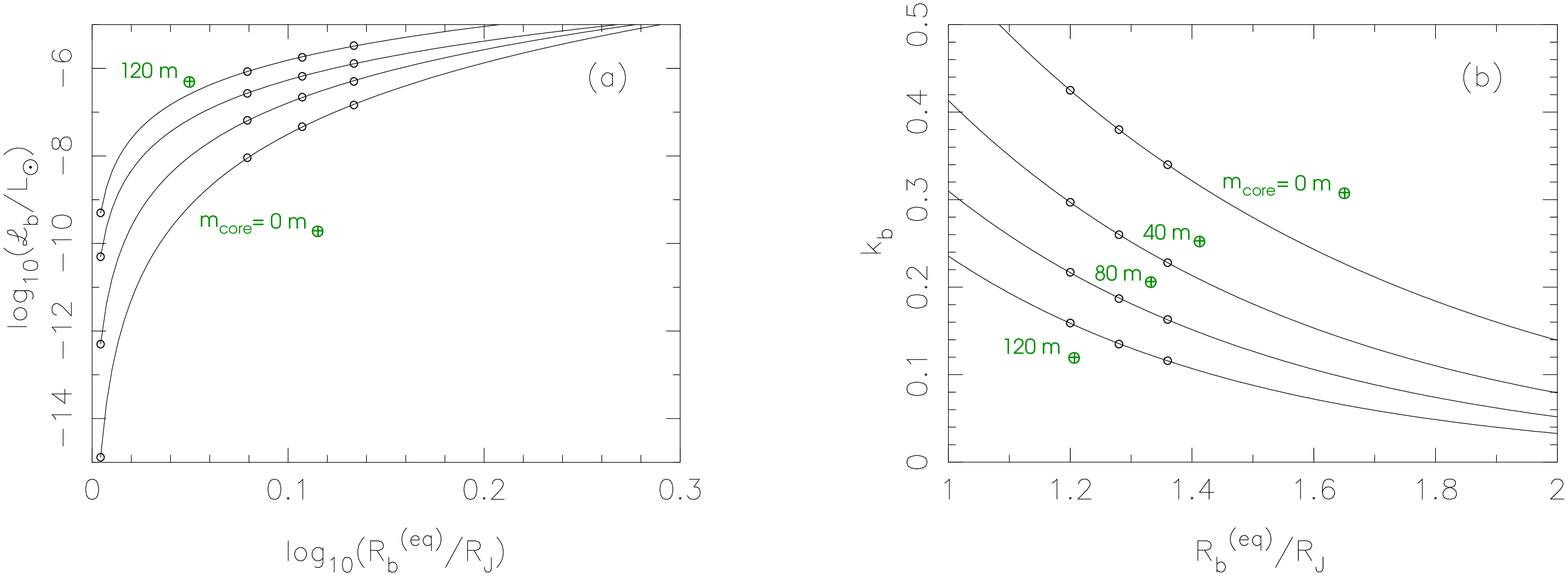}
\caption{Fitting functions for equations \rn{LR} and \rn{kb} for various values of the core mass of planet b.
(a): Log-log relations between the 
tidal power required to maintain the equilibrium
radius $R_b^{(eq)}/R_J$ given the cooling rate $-{\cal L}_b$ for such a structure, for core masses $m_{core}=0$, 40, 80 and $120 \,M_\oplus$.
Data is from \citet{batygin} (circles) except for the endpoints corresponding to $R_b^{(eq)}=1.01R_J$ which are
chosen for best fit and so that the curves have the property that $\lim_{R_b\rightarrow R_J}{\cal L}_b=0$.
(b): Exponential fitting laws for $k_b$ as a function of $R_b^{(eq)}/R_J$ for the same selection of core masses as in (a).
}
\label{LRkb}
\end{figure}

For each value of the core mass of planet b, $m_{core}$, \citet{batygin} provide three sets of values
of $(R_b^{(eq)}/R_J,\dot E_{tide})$, where $\dot E_{tide}$ is the tidal power required to
maintain the radius at $R_b^{(eq)}$ given the cooling rate $-{\cal L}_b$ for such a structure.\footnote{Note 
that stellar insolation is also included in their calculations so our 
values for $e_b^{(av)}$ below are slightly overestimated.}
The three values of $R_b^{(eq)}/R_J$ correspond to the best-fit 
observed value $\pm1\sigma$, these being (1.20,1.28,1.36). Desirable properties of the relationship between 
${\cal L}_b=\dot E_{tide}$ and $R_b^{(eq)}/R_J$ are
$\lim_{R_b\rightarrow R_J}{\cal L}_b= 0$ and
$d^2 (\log_{10}{\cal L}_b)/d(\log R_b)^2<0$ (see Figure~3 of \citet{BLM}).
A function with these properties is
\be
{\cal L}_b=c_n\left[\log_{10}(R_b^{(eq)}/R_J)\right]^{b_n} L_\odot,
\label{LR}
\ee
with least-squares parameters $b_n$ and $c_n$ depending on
$m_{core}$, as well as the choice for ${\cal L}_b(R_b^{(eq)}/R_J=1.01)/L_\odot\equiv \lambda_n$
(see Figure~\ref{LRkb}(a)).
These are listed in Table~\ref{lum}.
\begin{table*}
 \centering
 \begin{minipage}{140mm}
  \caption{Data for cooling and Love number fitting laws.}
  \label{lum}
  \begin{tabular}{ccccccc}
  \hline
$n$ & $m_{core}/M_\oplus$ & $b_n$ &  $\log_{10}(c_n)$ & $\lambda_n$
& $d_n$ & $p_n$  \\
\hline
0 & 0 & 5.41 & $-2.09$ & $-14.3$ & 2.27 & 1.39  \\
1 &40 & 4.04 & $-2.75$ & $-12.3$  & 2.16 & 1.65  \\
2 & 80 & 2.96 & $-3.31$ & $-10.3$  & 1.85 & 1.79  \\
3 & 120 & 2.56 & $-3.25$ & $-19.3$  & 1.69 & 1.97  \\
\hline
\end{tabular}
\end{minipage}
\end{table*}
Also listed are least-squares exponential-fit parameters for the Love number for each core mass such that
\be
k_b=d_n\,{\rm e}^{-p_n(R_b^{(eq)}/R_J)}.
\label{kb}
\ee
Fitting laws \rn{LR} and \rn{kb} are plotted in Figure~\ref{LRkb}.
Following \citet{ML} Section 4, the rate of change of the radius of planet b is given by
\be
\frac{\dot R_b}{R_b}=\frac{-{\cal L}_b+\dot E_{tide}}{\ff{1}{2}Gm_b^2/R_b+\alpha_b m_b R_b^2n_b^2}
\simeq
\frac{-{\cal L}_b/|E_b|+\tau_a^{-1}}{(m_b/m_*)(a_b/R_b)},
\label{RdotR}
\ee
where $\alpha_b$ is the moment of inertia coefficient of planet b\footnote{Formally $\alpha_b$ is also a function
of $R_b$, however, we take it to be constant at 0.26 in the numerical integrations presented in Section~\ref{origin},
corresponding to the moment of inertia coefficient of a polytrope of index 1.}
and $E_b=-\ff{1}{2}Gm_*m_b/a_b$ is the orbital binding energy, while
to ${\cal O}(e_b^2)$, ${\cal O}(\Omega_e^2/n_b^2)$ and ${\cal O}(\Omega_q^2/n_b^2)$, 
the orbit-averaged expression for $E_{tide}$ is given by\footnote{Note that equation
(73) in \citet{ML} should read $\langle\dot E_{tide}\rangle$ not $\langle\dot E_{tot}\rangle$. Note
also that their $k_2$ refers to the apsidal motion constant which is half the Love number.}
\be
\langle\dot E_{tide}\rangle=\ff{1}{2}(\mu_b a_b^2n_b^3)\left(\frac{k_b}{Q_b}\right)\left(\frac{m_*}{m_b}\right)
\left(\frac{R_b}{a_b}\right)^5\left[\left(\frac{\Omega_e}{n_b}\right)^2+\left(\frac{\Omega_q}{n_b}\right)^2
+21e_b^2\right].
\label{Etide}
\ee
Here $\Omega_e=\bOmega_b\cdot\hat{\bf e}_b$ and $\Omega_q=\bOmega_b\cdot\hat{\bf q}_b$,
where $\bOmega_b$ is the spin vector of planet b and $\hat{\bf q}_b=\hat{\bf h}_b\times\hat{\bf e}_b$, with
$\hat{\bf e}_b$ and $\hat{\bf h}_b$ unit vectors in the direction of periastron and 
planet b's orbital angular momentum per unit mass respectively. Note that \rn{Etide} assumes $\bOmega_b\cdot\hat{\bf h}_b=n_b$,
that is, planet b is synchronously rotating with its orbit.
The quantities $\Omega_e$ and $\Omega_q$ can remain non-zero for much longer
than the tidal damping timescale if $\bOmega_*\times{\bf h}_b\neq {\bf 0}$
and/or ${\bf h}_b\times{\bf h}_c\neq {\bf 0}$, where $\bOmega_*$ is the star's spin vector and
${\bf h}_c$ is the orbital angular momentum per unit mass of planet c.
The terms involving $\Omega_e$ and $\Omega_q$ in \rn{Etide} are associated with the so-called {\it obliquity tide} 
\citep{fabrycky2};
for the HAT-P-13 system they contribute less than 1\% to $\langle\dot E_{tide}\rangle$ for all
relative inclinations studied numerically, and will from hereon be ignored.
The planet's radius will increase or decrease according to whether $\dot E_{tide}$ is greater
or less than ${\cal L}_b$, at a rate enhanced by the factor $(m_*/m_b)(R_b/a_b)$ which is 21.5 
for the HAT-P-13 system. Once the system achieves equilibrium (ie, once $\dot E_{tide}={\cal L}_b$),
the radius of planet b will shrink on the timescale $\tau_a$, that is, the system will remain in the equilibrium state
appropriate for the current value of $a_b$. An example illustrating this behaviour is shown in
panel (i) of Figure~\ref{model1}.

\begin{figure}
\centering
\includegraphics[width=170mm]{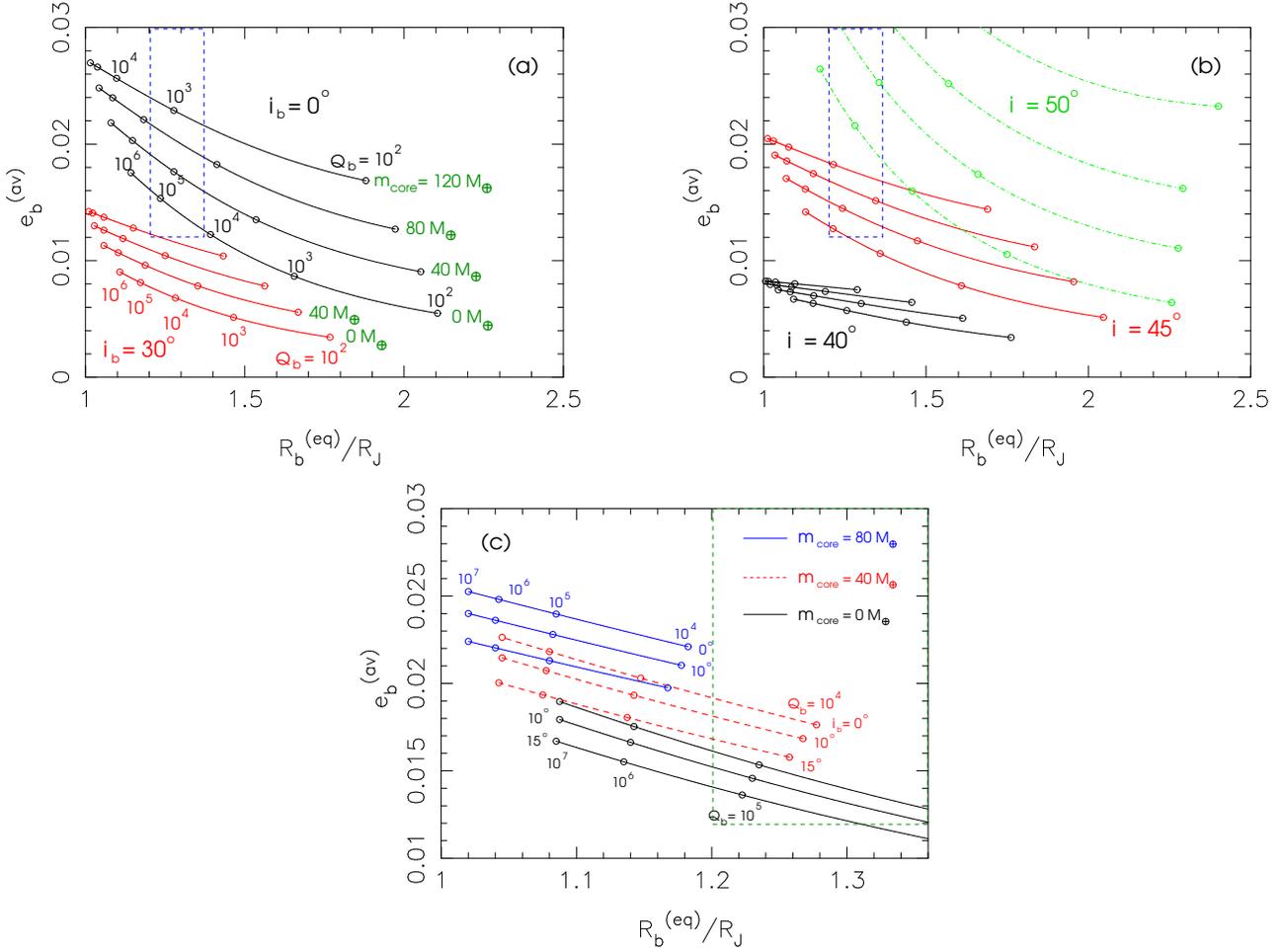}
\caption{The relationship between $e_b^{(av)}$, the equilibrium radius of planet b, its $Q$-value and its core mass
for the HAT-P-13 system {\it (note different scales on panel (c)).}
(a): Sets of curves for $i_b=0$ (black) and $i_b=30^\dg$ (red),
with the core mass increasing from bottom ($m_{core}=0$) to top ($m_{core}=120 M_\oplus$),
and $m_3=m_3^{min}/\cos i_b$.
The open circles correspond from right to left to $Q_b=10^2$, $10^3$, $10^4$,
$10^5$ and $10^6$; notice how they tend to bunch up as $Q_b$ increases and $R_b^{(eq)}\rightarrow 1$.
The blue dashed box indicates the current $1\sigma$ range of values
of $e_b^{(av)}$ and $R_b^{(eq)}$.
(b): Sets of curves for $i_b=40$ (black), $i_b=45^\dg$ (red) and $i_b=50^\dg$ (green dot-dashed),
with core masses and $Q$-values the same as in panel (a).
Since $i_b>i_b^{(c_1)}$, the expression used for the eccentricity of planet b is 
$e_b^{(av)}=\ff{1}{2}(e_b^{min}+e_b^{max})$, where $e_b^{min}$ and $e_b^{max}$
are taken from Table~\ref{ebminmax}.
Given the lower bound
placed on $Q_b$ by $\tau_e$ (Section~\ref{coplanar}), panels (a) and (b) suggest that for mutually prograde systems,
the orbits of planets b and c
are likely to be near coplanar, or have mutual inclinations between around $45^\dg$ and $50^\dg$.
Lower rather than higher core masses are also favoured.
(c): A alternative view. Sets of curves for $m_{core}=0\,M_\oplus$ (black), 
$40\,M_\oplus$ (red dashed) and $80\,M_\oplus$ (blue)
for (from top to bottom of each set) $0^\dg$, $10^\dg$ and $15^\dg$, and (circles from right to left along each curve)
$Q_b=10^4$, $10^5$, $10^6$ and $10^7$. The green dashed box indicates
current $1\sigma$ range of values
of $e_b^{(av)}$ and $R_b^{(eq)}$.
}
\label{eReq}
\end{figure}

For fixed $Q_b$ and $m_{core}$, \rn{LR} and \rn{Etide} may be equated to give a relationship between
$e_b$ and $R_b^{(eq)}$ (using \rn{kb} to express $k_b$ in terms of $R_b$). Equation \rn{equilinc} provides a second relationship between these two variables for
fixed $i_b$, and
combining the two gives one between $R_b^{(eq)}$ and $Q_b$. Solving this for $R_b^{(eq)}$
for $10^2\leq Q_b\leq 10^6$ and substituting the resulting values into \rn{equilinc} allows us to plot
$e_b^{(av)}$ as a function of $R_b^{(eq)}$ and $Q_b$ for each value of $m_{core}/M_\oplus$ presented
in \citet{batygin}. Such curves are plotted in panel (a) of Figure~\ref{eReq}
for $i_b=0$ and $30^\dg$, and panel (b) for $i_b=40^\dg$, $45^\dg$ and $50^\dg$.
Also plotted (blue-dashed box) is the current $1\sigma$ range of values
of $e_b^{(av)}$ and $R_b^{(eq)}$. 
Since $i_b>i_b^{(c_1)}$ for the curves in panel (b), the expression used for b's eccentricity is 
$e_b^{(av)}=\ff{1}{2}(e_b^{min}+e_b^{max})$, where $e_b^{min}$ and $e_b^{max}$
are taken from Table~\ref{ebminmax}. Given $\tau_e\propto \tau_a\propto e_b^{-2}$ (Section~\ref{coplanar}),
we were able to confirm numerically that using $e_b^{(av)}$ rather than $e_b^{max}$ gives a good estimate for the
decay timescale of $e_b$. This is in contrast to systems with high eccentricities (see Section~\ref{high}).

Figure~\ref{eReq}(c) presents an alternative view for low inclinations, 
with each set of curves corresponding to a different core mass.
Given the lower bound
placed on $Q_b$ by $\tau_e$ (Section~\ref{coplanar}), consistent with the fact that
higher rather than lower values of $Q_b$ are suggested by
the orbital parameters of short-period planets \citep{wu03}, Figure~\ref{eReq} suggests that 
the orbits of planets b and c
are likely to be either near coplanar (mutually prograde or retrograde; see Section~\ref{retrograd}), 
or have mutual inclinations between around $45^\dg$ and $50^\dg$ (or between $130^\dg$ and $135^\dg$). 
However, a consistency argument due to Daniel Fabrycky (private communication) rules out the
$45-50^\dg$ and near retrograde coplaner cases. The steps in the argument are as follows, where we take 
the mutual inclination of orbits b and c, $i_{bc}$, to be $50^\dg$
for definiteness:
\ben
\item
$\omega_b-\omega_c\equiv\Delta\omega=4\pm 46=-42$ - $50^\dg$;
\item
The angle $\eta=\omega_b-\omega_c+\Omega_b-\Omega_c$ librates around $180^\dg$
for orbits with $i_{bc}=50^\dg$, with a libration amplitude of around $50^\dg$.
Thus from (i),
$\Omega_b-\Omega_c\equiv \Delta\Omega=80-272^\dg$ so that $-1\leq\cos\Delta\Omega\leq 0.17$;
\item
Measuring inclinations with respect to the line of sight (as opposed to the invariable
plane normal), $i_b=83.4^\dg\pm 0.6$;
\item
$\cos i_{bc}=\sin i_b\sin i_c\cos(\Omega_b-\Omega_c)+\cos i_b\cos i_c\simeq \sin i_c\cos\Delta\Omega$;
\item
Since $0\leq\sin i_c\leq 1$, $\cos i_{bc}\leq \cos\Delta\Omega\leq 1$ from (iv). But since $\cos i_{bc}=\cos 50^\dg=0.64$,
this contradicts (ii). Thus a mutual inclination of  $50^\dg$ is inconsistent with observations.
\een
The argument against coplanar retrograde systems follows similarly, with $\cos i_{bc}\simeq -1$ and
the librating angle
$\zeta=\omega_b+\omega_c-\Delta\Omega=0$ so that $\Delta\Omega=-48$ - $44^\dg$ with 
$\cos\Delta\Omega>0$, making $\sin i_c<0$.

Figure~\ref{eReq} also suggests that lower rather than higher core masses are favoured.
Note that for inclined systems for which $i_b<i_b^{(c_1)}$, $\tau_e$ can be estimated
by replacing $\Delta_0$ with $\Delta$ in \rn{taue}, where $\Delta$ is given by \rn{Wq}.
Thus since $\Delta\leq\Delta_0$, the lower bound for $Q_b$ given by $\tau_e$ for coplanar systems represents
a lower bound for higher inclinations in this range.

More accurate measurements of $e_b$ and $R_b$ will allow refinement of the statements above.
Note that a small value of $i_b$ does {\it not} imply that a Rossiter-McLaughlin measurement of the stellar obliquity will result
in a small value; its maximum depends not only on the mutual inclination of the two orbits, but also on the stellar
obliquity {\it relative to orbit c} (see Section~\ref{origin}).

\subsection{Timescale for decay of the mutual inclination in a relaxed system}\label{incdec}

\begin{figure}
\centering
\includegraphics[width=170mm]{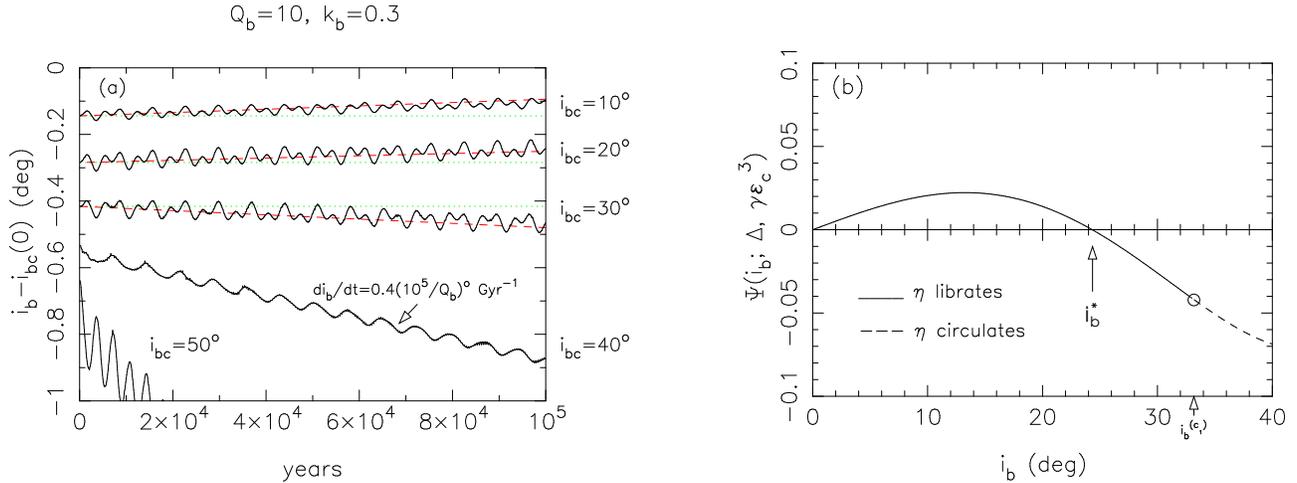}
\caption{(a) Decay (or increase) of $i_b$ for $Q=10$ and $k_b=0.3$,
where $i_b$ is measured relative to the invariable plane and
$i_{bc}(0)$ is the initial mutual inclination. Solid black curves represent numerical integrations, red dashed curves
give the trend according to \rn{didtav} and green dotted curves are horizontal reference curves to guide the eye.
Trend curves are not given for $i_{bc}(0)=40^\dg$ and $50^\dg$, systems for which the theory is no longer valid.
Note that for the cases $i_{bc}(0)=40^\dg$ and $50^\dg$
the oscillatory behaviour is dominated by the quadrupole contribution to $di_b/dt$ (which is proportional
to $\sin i_c$ and has frequency $\dot\Omega_b-\dot\Omega_c$) while for $i_b(0)=10^\dg$, $20^\dg$ and $30^\dg$ the 
octopole contribution is evident (see equation \rn{ip}).
(b) The function $\Psi(i_b;\Delta,\gamma\varepsilon_c^3)$ defined in equation \rn{Psi}.
Note that $\Psi\gl 0$ for $i_b\lg i_b^*$, consistent with numerical solutions.
}
\label{incdecay}
\end{figure}

Taking $i_b$ as a proxy for the mutual inclination (since we are assuming most of the angular momentum
of the system resides in the outer orbit), we can use \rn{ip} to estimate a timescale for the decay of the 
mutual inclination of a relaxed system. Figure~\ref{incdecay}(a)
shows $10^5$ years of evolution of $i_b$ for the systems shown in Figure~\ref{eeq4}, for which $Q_b=10$
and $k_b=0.3$. In order
to clearly demonstrate long-term trends, we have plotted
the quantity $i_b-i_{bc}(0)$, where $i_b$ is measured {\it relative to the invariable plane} and
$i_{bc}(0)$ is the initial mutual inclination.
In particular, note that the general trend is positive for $i_b=10^\dg$ and $20^\dg$, while for 
$i_b=30^\dg$, $i_b=40^\dg$ and $50^\dg$ it is negative. For systems for which
$\eta$ librates (here, $i_b=10^\dg$, $20^\dg$ and $30^\dg$), this can be understood as follows.

Consider equation~\rn{ip} for the rate of change of $i_b$. The current slope of the trend may be determined
by taking a time average of $di_b/dt$ holding $e_c$, $i_b$ and $a_b/a_c$ constant.
Replacing $e_b$ and $\sin\eta$ by their limit-cycle counterparts
$e_b^{(lc)}$ and $\eta^{(lc)}$ respectively (equation~\rn{eA}),
and recalling that $\varpi_b+\varpi_c-2\Omega_b=2\omega_b-\eta$, 
we obtain
\bea
\left<\frac{di_b}{dt}\right>&\equiv&
\lim_{T\rightarrow\infty}\frac{1}{T}\int_0^T\frac{di_b}{dt}=
-\frac{15}{32}n_b\left(\frac{m_c}{m_*}\right)\left(\frac{a_b}{a_c}\right)^4
\varepsilon_c^{-5}e_c\, e_b^{(av)} \eta^{(av)}\sin 2i_b \left[f_4(i_b)-\ff{1}{2}{\cal A}_{lc}^*h_4(i_b)\right]\next
&=&\frac{25}{32}\left(\frac{a_b}{a_c}\right)^2\,e_c^2\varepsilon_c^{-4}
\Psi(i_b;\Delta,\gamma\varepsilon_c^3)\,\cdot \tau_{circ}^{-1}\next
&\equiv&(i_b-i_b^*)\,\tau_i^{-1},
\label{didtav}
\eea
where 
\be
\Psi(i_b;\Delta,\gamma\varepsilon_c^3)=f_2(i_b)\sin(2i_b)\left[f_4(i_b)-\ff{1}{2}{\cal A}_{lc}^*(i_b)h_4(i_c)\right]\Delta^{-2}
\label{Psi}
\ee
with $\Delta$ defined in \rn{Wq}, $i_b^*$ is the first root of $\Psi$
and $\tau_i$ is time it takes for the system to relax to $i_b=i_b^*$.
For $i_b=30^\dg$, $\tau_i\simeq 80$~Gyr. 
The function $\Psi(i_b;\Delta,\gamma\varepsilon_c^3)$ is plotted in Figure~\ref{incdecay}(b)
as a function of $i_b$ for HAT-P-13 system parameters with $k_b=0.3$. We see that
$\Psi> 0$ for $i_b<i_b^*\simeq 24^\dg$, that is, $i_b$ actually {\it increases} for this range of inclinations,
while for greater inclinations $i_b$ decreases. 
The red dashed lines in Figure~\ref{incdecay}(a) have slopes equal to $\langle di_b/dt\rangle$, and are in good agreement
with the general trend of the numerical solutions. Also shown are numerical solutions for
$i_b=40^\dg$ and $50^\dg$. The slope of the trend for $i_b=40^\dg$ is $0.4(10^5/Q_b)^\dg\,{\rm Gyr}^{-1}$,
while that for  $i_b=50^\dg$ is a factor of three higher.
Thus the timescale for the decay of the mutual inclination of HAT-P-13-like systems for which $i_b<50^\dg$
is considerably longer than the age of the system for reasonable values of $Q_b$.

A similar analysis can be done for retrograde systems, and
we can conclude generally that
{\it inclined systems cannot relax to the coplanar prograde or retrograde state as long as} $e_b^{(av)}>0$,
and that they relax to a mutual inclination given by one of the roots of $\Psi(i_b)=0$ (or $\Psi(\pi-i_b)=0$ for retrograde)
as long as $\tau_i<\tau_c$ and $\tau_i<\tau_a$. 
Note that the {\it mutual} inclination changes by the same amount as $i_b$, and that
\rn{didtav} does not apply to systems with
mutual inclinations greater than $i_b^{(c_1)}$ for which it represents a lower bound.

\section{A scenario for the origin of HAT-P-13-like systems}\label{origin}

The high eccentricity of HAT-P-13c suggests either a violent scattering history, or Kozai-type interactions
with another planet or star, or that it was formed through gravitational collapse; it seems unlikely that
such a high eccentricity could result from single planet-disk interactions alone (see, for example,
\citet{artymowicz}). Kozai forcing
would also affect planet b, and it is unclear without a detailed study
whether or not a suitable configuration exists which would not cause the rapid decay of its orbit.

Here we focus on the scattering scenario, introducing a third planet (``planet d'') which is ultimately
ejected from the system. The mass ratio of planet b to planet c is too low for c to have attained its high eccentricity
during a scattering event with b.
Three models are presented, all of which have identical initial conditions except for the eccentricity
of planet d, $e_d$,
which for model 2 differs from 1 by two parts in $10^5$ while for model 3 it differs by 0.05.
The outcomes are significantly different, with model 1 producing a value for $e_c$ almost identical to that for HAT-P-13c,
while models 2 and 3 produce lower values at 0.46 and 0.36. Of particular interest is the relative inclination of the
orbits of b and c
following the escape of d, and the accompanying stellar obliquity relative to planet b, both of which
differ significantly from model to model, as does the equilibrium value of $R_b$.

The initial configuration data are listed in Table~\ref{data},
\begin{table*}
\label{data}
 \begin{minipage}{175mm}
\caption{Data for models 1, 2 and 3 at $t=0$ and around time planet b first achieves equilibrium}
  \begin{tabular}{@{}llllllllllll@{}}
  \hline
$Q_b=40$ && \multicolumn{2}{c}{model 1\phantom{ccccc}} &&&  \multicolumn{2}{c}{model 2\phantom{ccccc}} && & 
 \multicolumn{2}{c}{model 3\phantom{ccccc}}   \\ \hline
$t$ (yr) && 0 & 1.5 Myr &&& 0 & 10 Myr &&&  0 & 35 Myr\footnote{Equilibrium has not yet been established.}  \\  \hline
&&&&&&&&&&&\\
$m_b\,(M_J)$ && 0.85 & &&& 0.85 & &&& 0.85 &  \\
&&&&&&&&&&&\\
$m_c\,(M_J)$ && 15.2 & &&& 15.2 & &&& 15.2 &  \\
&&&&&&&&&&&\\
$m_d\,(M_J)$ && 12 & &&& 12 & &&& 12 &  \\
&&&&&&&&&&&\\
&&&&&&&&&&&\\
$a_b\,(AU)$ && 0.043 & 0.0406 &&& 0.043 &0.0422 &&& 0.043 & 0.04255 \\
&&&&&&&&&&&\\
$a_c\,(AU)$ && 1.7 & 1.17 &&& 1.7 & 1.16 &&& 1.7 & 1.17  \\
&&&&&&&&&&&\\
$a_d\,(AU)$ && 3.097 & $\infty$&&& 3.097 & $\infty$ &&& 3.097 & $\infty$ \\
&&&&&&&&&&&\\
$R_b\,(R_J)$ && 1.5 & {\bf 2.34} &&& 1.5 & {\bf 1.44} &&& 1.5 & {\bf 1.14} \\
&&&&&&&&&&&\\
&&&&&&&&&&&\\
$e_b$ && 0.005 & {\bf 0.007} 
&&& 0.005 & {\bf 0.004} &&& 0.005 & {\bf 0.001}  \\
&&&&&&&&&&&\\
$e_c$ && 0.1 & {\bf 0.68} &&& 0.1 & {\bf 0.45} &&& 0.1 & {\bf 0.36}  \\
&&&&&&&&&&&\\
$e_d$ && 0.05 & 1.013 &&& 0.05002 & 1.05 &&& 0.1 & 1.01  \\
&&&&&&&&&&&\\
&&&&&&&&&&&\\
$\psi_{*b}$ && $0^\dg$ & ${\bf 3.5-18}^\dg$ &&& $0^\dg$ &  ${\bf 3-40}^\dg$  &&& $0^\dg$ &  ${\bf 23-53}^\dg$   \\
&&&&&&&&&&&\\
$\psi_{*c}\simeq\theta_*$ && $15^\dg$ &  ${\bf 7-8}^\dg$  &&& $15^\dg$ &  ${\bf 18-21}^\dg$ &&& $15^\dg$ &  ${\bf 12-19}^\dg$  \\
&&&&&&&&&&&\\
&&&&&&&&&&&\\
$i_b^{(i)}$ && $10^\dg$ & $6-14.5^\dg$&&& $10^\dg$ & $8-32^\dg$ &&& $10^\dg$ & $30-38^\dg$ \\
&&&&&&&&&&&\\
$i_c^{(i)}$ && $5^\dg$ & $4^\dg$ &&& $5^\dg$ & $11.3^\dg$ &&& $5^\dg$ & $4^\dg$ \\
&&&&&&&&&&&\\
$i_d^{(i)}$ && $5^\dg$ & $2^\dg$ &&& $5^\dg$ & $6.2^\dg$ &&& $5^\dg$ & $2.3^\dg$  \\
&&&&&&&&&&&\\
&&&&&&&&&&&\\
$i_b^{(f)}\simeq i_{bc}$ && $15^\dg$ & ${\bf 10.2-10.6}^\dg$  &&& $15^\dg$ & ${\bf 19.5-21}^\dg$  &&& $15^\dg$ & ${\bf 34-35.5}^\dg$  \\
&&&&&&&&&&&\\
$i_c^{(f)}$ && &${\bf 0.2}^\dg$  & &&& ${\bf 0.2}^\dg$ & &&&  ${\bf 0.35}^\dg$    \\
&&&&&&&&&&&\\
$\theta_{IP}$ && &$4^\dg$  & &&& $11.2^\dg$ & &&&  $3.8^\dg$    \\
&&&&&&&&&&&\\
\hline
\end{tabular}
\end{minipage}
\end{table*}
together with those at the time planet b first achieves an equilibrium radius, that is an equilibrium between
the rate at which tidal energy is injected and the rate at which the planet can cool.
Data of particular significance are highlighted in bold.
Inclinations with superscripts $(i)$ and $(f)$ are measured with respect to the original and final invariable planes
respectively, the former including planet d and the latter not. The quantity $\theta_{IP}$
is the angle between the invariable planes before and after the escape of planet d.
Again we use the averaged code of \citet{ML}, this time without suppressing the
evolution of the radius of planet b which is evolved according to the scheme discussed
here in Section~\ref{seceReq}
with $m_{core}=80M_\oplus$.
Note that this code involves averaging over the innermost orbit only;
the two outer orbits are integrated directly and are therefore susceptible to instability through the 
overlap of mean-motion resonances \citep{stability}.
While the mass of planet b is taken as that of HAT-P-13b, 
the mass of planet c is taken as the {\it minimum} mass of HAT-P-13c,
that is, it is not scaled by $\cos i_b$. The mass of planet d is taken to be
similar but less than that of planet c, ensuring that the probability that planet c is left with a high
eccentricity is significant.
Initial values of the eccentricities and inclinations are chosen to be consistent with formation
and subsequent migration in a protoplanetary disk, and the period ratio (which is similar to that
for Jupiter and Saturn) puts the system just inside the
stability boundary of a system with these masses and eccentricities (see, for example, Figure 9(a)
of \citet{stability}), consistent with having just emerged from the protection of the disk.\footnote{The
disk will still be present at this stage, but its density will not be high enough to suppress eccentricity
growth and prevent the overlap
of neighbouring mean-motion resonances.}
Moreover, the initial values of $a_c$ and the semimajor axis of planet d, $a_d$, are such that 
the energy needed for escape of planet d and provided by the orbit of planet c reduces
$a_c$ to a value similar to that of HAT-P-13c.
The initial radius of planet b is $1.5R_J$, consistent with the radius of a young planet
recently arrived at its present location (see Figure 1 of \citet{BLM}; disk lifetimes suggest that such a 
planet would probably have a higher radius).
The initial mean longitudes and orientation angles of planets b and c are specified relative to the outer orbit;
these are 
$\lambda_{bd}=0^\dg$,
$i_{bd}=5^\dg$, $\omega_{bd}=0^\dg$, $\Omega_{bd}=0^\dg$ and 
$\lambda_{cd}=180^\dg$, $i_{cd}=10^\dg$, $\omega_{cd}=0^\dg$, $\Omega_{cd}=180^\dg$ 
respectively and are such that longitudes are measured with respect to the periastron direction of orbit d.

\begin{figure}
\centering
\includegraphics[width=160mm]{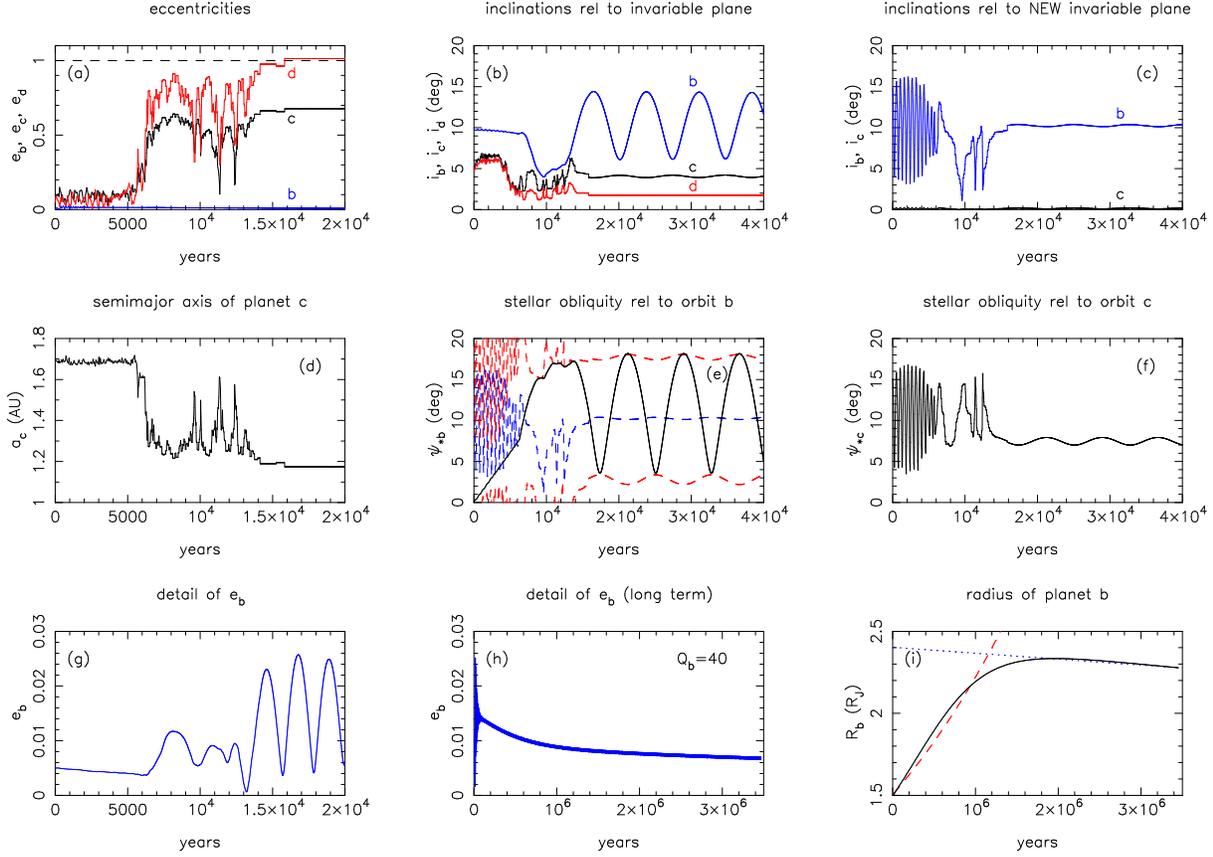}
\caption{A chaotic origin for the HAT-P-13 system {\it (note different timescales for each panel)}. 
At time $t=0$ three planets are present, with
migration leaving planet b 0.043 AU from the star with a radius of $1.5 R_J$ and a core of mass $80M_\oplus$, 
and planets c and d with period ratio 2.46.
While the presence of an outer disk has previously 
limited the variation of the eccentricities thereby protecting
the system against instability, by $t=0$ the disk surface density has reduced sufficiently to allow the
system to become unstable. Panel (a) shows all three eccentricities, each remaining moderate for the 
first 6000 years.
After 16,000 years planet d escapes the system. 
Following escape, the orbit of planet b precesses around the
{\it new} invariable plane whose normal is approximately parallel to that of planet c (since with
d removed from the system it
now contains 99\% of the total angular momentum; compare panels (b) and (c)).
The inclinations of the three orbits to the original invariable plane are modest initially, and they remain so during and after
the scattering process. 
Following the escape of planet d, the stellar obliquity relative to b's orbit (panel (e),
black solid curve) oscillates about a mean equal to $i_b$ (blue dashed curve) and with an
amplitude equal to $\psi_{*c}$, the stellar obliquity relative to c's orbit (red dashed curves; see also
panel (f) and Figure~\ref{obliquity}). Panel (g) shows the early behaviour of $e_b$; before the scatter
planet b is effectively decoupled from the rest of the system and its eccentricity monotonically
decreases on the tidal circularization timescale, while after the scatter it is governed by planet c.
An artificially low $Q$-value of 40 is used for planet b, and its radius and Love number are evolved
according to \rn{LR} and \rn{kb} respectively ($k_b(0)=0.13$ for $R_b(0)=1.5R_J$ and $m_{core}=80M_\oplus$).
Panels (h) and (i) show the long-term evolution of $e_b$ and $R_b$
respectively. The system evolves to a limit cycle after about 0.12 Myr,
with $e_b^{(av)}$ decreasing as
$R_b$ increases (given the dependence of $e_b^{(av)}$ on $R_b$ via $\gamma_b^{tide}\propto (R_b/a_b)^5$)
until an equilibrium value of $R_b$, $R_b^{(eq)}=2.33R_J$, 
is reached after 1.6 Myr. 
This value of $R_b^{(eq)}$ corresponds to $e_b=0.007$ and agrees favorably 
with the theoretical values $R_b^{(eq)}=2.12R_J$
and $e_b=0.008$ obtained using
the procedure described in Section~\ref{seceReq}.
Note that more realistic (ie, higher) values for $Q_b$ would result in higher equilibrium values of $e_b$ 
and lower values of $R_b^{(eq)}$ (see Figure~\ref{eReq}).
During the 3.5 Myr integration, the semimajor axis decreases from 0.04300 to 0.04055 AU, giving
an orbital decay timescale $\tau_a=a_b/\dot a_b$ of $158 (Q_b/10^5)$ Gyr.
Also shown in panel (i) are curves $R_b(t)=R_b(0){\rm exp}(t/\tau_+)$ (red dashed curve) and
$R_b(t)=R_b(t_*){\rm exp}[-(t-t_*)/\tau_a]$ (blue dotted curve),
where $\tau_+=(m_b/m_*)(a_b/R_b)\tau_a$ and $t_*=3.5$ Myr. These curves demonstrate that initially
the rate of change of the radius is dominated by tidal heating, that is, the cooling term
contributes very little (see equation~\rn{RdotR}), 
and once equilibrium is reached it is
continually reestablished as $a_b$ decreases on the timescale $\tau_a$.
}
\label{model1}
\end{figure}

\begin{figure}
\centering
\includegraphics[width=160mm]{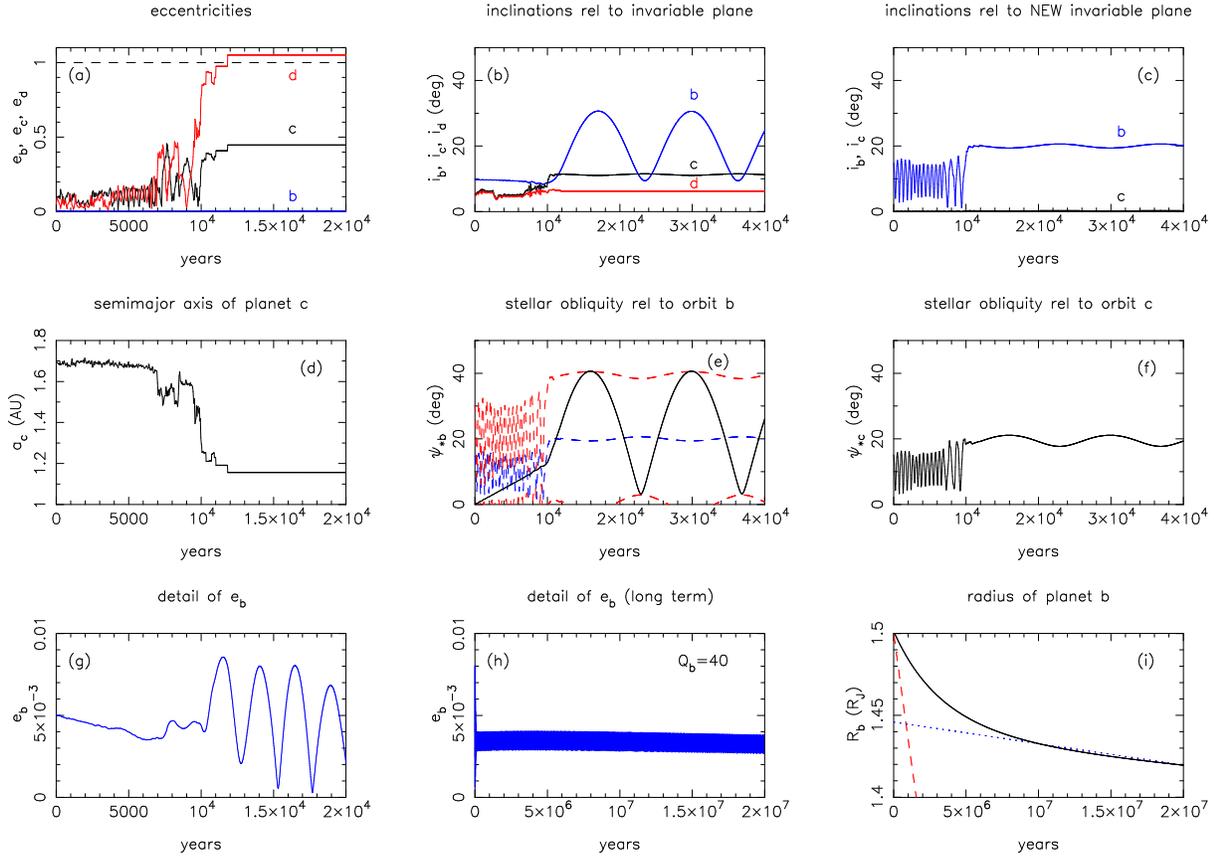}
\caption{A different outcome for the HAT-P-13 system {\it (note different timescales for each panel)}. 
Initial conditions are the same as for model 1 but with a difference in $e_d$ of $2\times 10^{-5}$.
Following the scatter of planet d from the system, $e_c=0.45$ (panel (a)), resulting in a smaller value
of $e_b^{(av)}$ (panel (g)). This time the mutual inclination reaches $20^\dg$, as does
the stellar obliquity relative to c's orbit (panels (c) and (f)). This results in a maximum stellar obliquity
relative to b's orbit of $i_b+\psi_{*c}=40^\dg$ 
and a minimum of $3^\dg>i_b-\psi_{*c}$ (panel (e); the variation of $\psi_{*b}$ is not a simple sinusoid
when its mimimum is near zero, similar to the behaviour of eccentricity in the same circumstances).
Note that the precession timescale is longer than that of model 1 by a factor 
$(\varepsilon_c^{(2)}/\varepsilon_c^{(1)})^3$, where the superscripts refers to the model numbers
(see equation~\rn{WOmega}).
The long-term evolution of $e_b$ (panel (g)) is a limit cycle for which $e_b^{(av)}$ {\it increases} slightly
for the first 10 Myr or so, corresponding to the relatively rapid (but still very slow) decrease of $R_b$ (panel (h)),
then decreases once equilibrium is established.
The theoretical prediction according to Section~\ref{seceReq} gives $R_b^{(eq)}=1.42R_J$ and $e_b^{(av)}=0.0037$
when $a_b=0.0426$.
During the 20 Myr integration, the semimajor axis decreases from 0.0430 to 0.0422 AU, giving
an orbital decay timescale $\tau_a=2590 (Q_b/10^5)$ Gyr.
Also shown in panel (i) are curves $R_b(t)=R_b(0){\rm exp}(-t/\tau_-)$ (red dashed curve) and
$R_b(t)=R_b(t_*){\rm exp}[-(t-t_*)/\tau_a]$ (blue dotted curve),
where $\tau_-=(m_b/m_*)(a_b/R_b)|E_b|/{\cal L}_b(R_b(0))$ and $t_*=20$ Myr. These curves demonstrate that 
the rate of change of the radius is never dominated either by heating or cooling (see equation~\ref{RdotR}), 
being close to equilibrium initially. After around 10 Myr equilibrium is established 
and $R_b$ decreases on the timescale $\tau_a$.
}
\label{model2}
\end{figure}

\begin{figure}
\centering
\includegraphics[width=160mm]{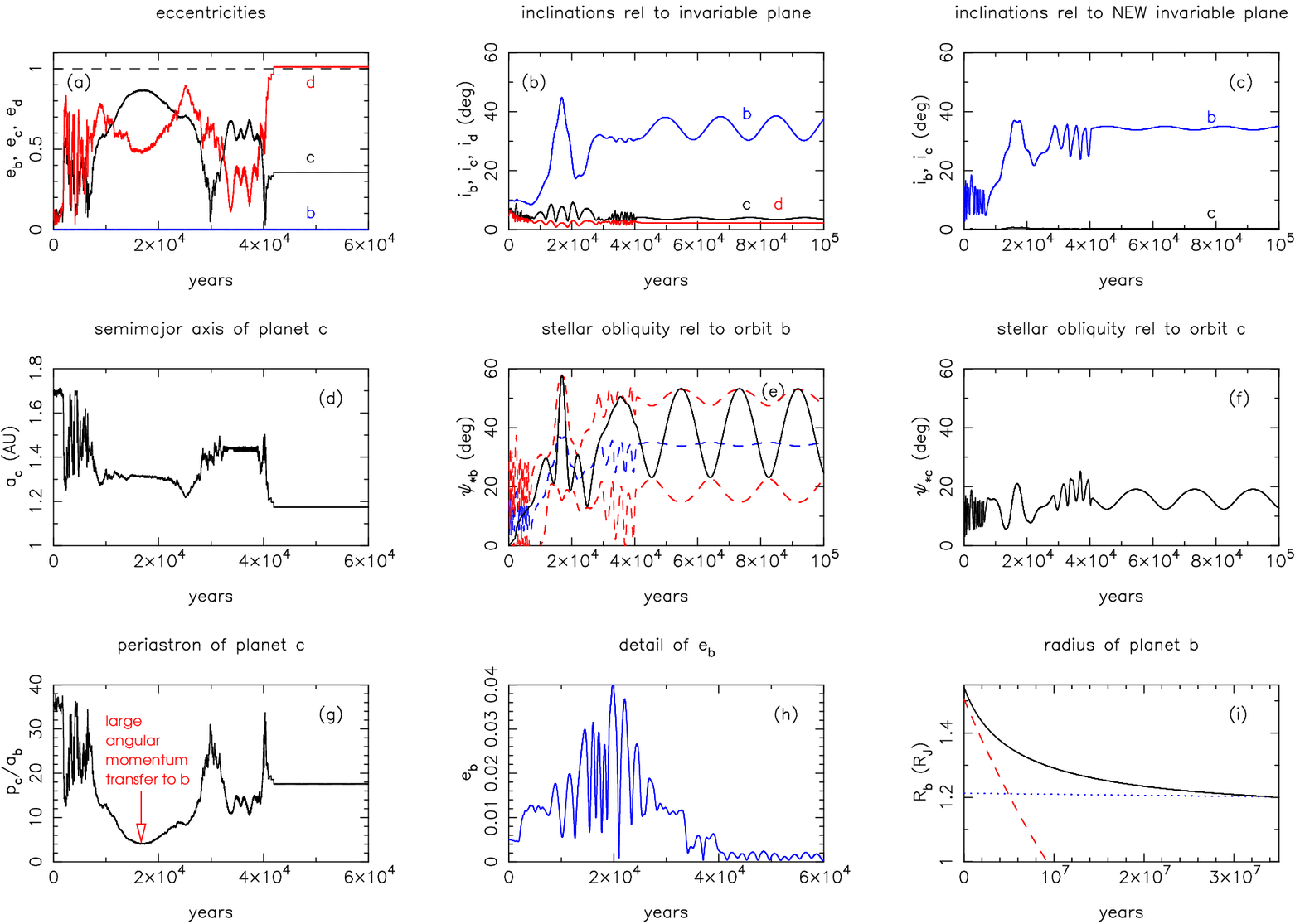}
\caption{High stellar obliquity due to a close encounter between b and c {\it (note different timescales for each panel)}. 
Initial conditions are the same as for model 1 but with $e_d=0.1$.
This time the mutual inclination reaches $35^\dg$, while
the maximum stellar obliquity relative to c's orbit after the scatter is $19^\dg$
(panels (c) and (f)). This results in a maximum stellar obliquity
relative to b's orbit of $i_b+\psi_{*c}=53^\dg$ 
and a minimum of $i_b-\psi_{*c}=23^\dg$ (panel (e)).
Unlike models 1 and 2 where the stellar obliquity 
can be attributed almost entirely to angular momentum transfer between c and d (and the initial value of $\psi_{*c}$),
$i_b$ and $\psi_{*c}$ are significantly different, a result of a close encounter
between planets b and c around 17,000 yr when $e_c$ reaches 0.87 (see panel (g) in which
the periastron separation, $p_c$, reaches a minimum of $4a_b$). 
During this high-eccentricity phase, the torque from planet c tilts the orbit of planet b through $30^\dg$,
and while some of the associated angular momentum is subsequently returned to c's orbit, the mutual
inclination remains high following the escape of planet d.
Since the value of $e_c$ is relatively low at 0.36 after the escape (panel (a)), 
$e_b^{(av)}$ is a mere 0.001 (panel (h)), with a similarly
small amplitude in spite of the relatively high mutual inclination. As a consequence, tidal heating is weak
and the predicted equilibrium radius of planet b is only $1.14 R_J$ (recall $Q_b=40$). 
Panel (i) shows the evolution of $R_b$ together with curves $R_b(t)=R_b(0){\rm exp}(-t/\tau_-)$ (red dashed curve) and
$R_b(t)=R_b(t_*){\rm exp}[-(t-t_*)/\tau_a]$ (blue dotted curve),
where this time $t_*=35$ Myr. These curves demonstrate that initially
the rate of change of the radius is dominated by cooling (see equation~\ref{RdotR}), 
and does not appear to reach equilibrium during the time shown.
}
\label{model3}
\end{figure}

The time evolution of various quantities is plotted in Figures~\ref{model1}, \ref{model2} and \ref{model3},
with detailed descriptions and discussion provided in the captions. 
The choice of an initial period ratio for orbits c and d of around 5:2 affects possible outcomes
in the following ways (given that we wish to approximately reproduce the HAT-P-13 system).
In view of the fact that we wish to place the system near the edge of the stability boundary, the choice of $e_d(0)$ is
restricted by the fact that a value significantly higher would render the system far from the boundary and hence violently unstable.
Our choice of $m_d$ affects the choice of $a_c(0)$; a smaller value of $m_d$ requires a smaller value
of $a_c(0)$, however, it tends to produce smaller values of $e_c$
following escape of d and is less likely to produce high stellar obliquities (see model 3).
On the other hand, a heavier companion is more likely to eject planet c (an outcome also possible
for $m_d=12M_J$; here we restrict ourselves to systems in which the outer planet is ejected). 
A general detailed study is required to quantify possible outcomes of such a scattering scenario,
especially in the light of the recent discovery of two retrograde systems \citep{winn,anderson}.
For now our aim is to demonstrate its potential to produce a range of eccentricities,
mutual inclinations, stellar obliquities and planetary radii.

Of particular interest is the stellar obliquity relative to
the orbit of planet b, $\psi_{*b}$, a quantity
which can be measured directly (at least in sky projection; see \citet{fabryckywinn} for a discussion
of the statistical properties of this quantity). In the following section we outline the mechanics
of stellar obliquity in a two-planet system.

\subsection{Stellar obliquity in a two-planet system}\label{twoplanet}

The stellar obliquity relative to each planetary orbit in a system reflects conditions at the time of its formation.
In the scattering models presented in this section we assume zero stellar obliquity relative to planet b
initially, while the stellar obliquity relative to planet c is $15^\dg$. Table~\ref{data} lists the ranges for
$\psi_{*b}$ following the escape of planet d, demonstrating that significantly different outcomes are
possible from models with very similar initial conditions. The variation in $\psi_{*b}$ depends 
on two quantities: the angle between the stellar spin axis and the normal to the invariable plane, $\theta_*$, and
the angle between planet b's orbit normal and the normal to the invariable plane, $i_b$.
When $\psi_{*b}\neq 0$, the variable
torque on the spin bulge of the star from planet b results in nutation of the star's spin axis, that is, a variation
of $\theta_*$.
This can be quite significant; in model~3 it is $7^\dg$ compared with $1^\dg$
for model 1 (see Table~\ref{data}).\footnote{Note that 
the moment of inertia of the star is about three times that of b's orbit.}
Its average value, however, depends on the history of the system; 
if no planets have been ejected since the system's formation, $\theta_*$ is likely to be modest, 
while the escape of one or more planets 
can, depending on how much angular momentum the planets carry away, 
result in an invariable plane normal which points in a significantly different direction to the original
($\theta_{IP}$ in Table~\ref{data}), thereby affecting $\theta_*$.
\citet{barker} have shown that the decay timescale for $\psi_{*b}$ is around $\tau_a$, the
timescale for the decay of the orbit, so very little reduction in the average value of $\theta_*$ is expected over the lifetime
of the orbit of HAT-P-13b.

Regarding the relevant angles as spherical polar angles, 
$\psi_{*b}$ is given in terms of $\theta_*$, the spin axis node angle,
$\varphi_*$, as well as $i_b$ and $\Omega_b$, by
\be
\cos \psi_{*b}=\hat{\bOmega}_*\cdot\hat{\bf h}_b
=\sin\theta_*\sin i_b\cos(\varphi_*-\Omega_b)+\cos\theta_*\cos i_b.
\ee
Since the variations in $\theta_*$ and $i_b$ are small (but see next section),
$\psi_{*b}$ cycles approximately between $|i_b+\theta_*|$ and
$|i_b-\theta_*|$ as $\varphi_*-\Omega_b$ cycles between 0 and $2\pi$ over b's
precession cycle,\footnote{Note that      
the precession rate of the spin axis node
is more than 15 times slower than that of b's orbit for the examples in this section.}
or, since the orbit of planet c contains most of the angular momentum of the system so that $\theta_*\simeq\psi_{*c}$,
the variation is approximately $|i_b\pm\psi_{*c}|$.
In fact, since the nutation period of the star is equal to the precession period of planet b, extrema
of $\psi_{*b}$ coincide with extrema of $\psi_{*c}$  (see panel (e) of Figures~\ref{model1} to \ref{model3}).

A significant difference between the first two models and model 3 is that
before the escape of planet d, the latter suffers a significant transfer
of angular momentum between orbits c and b during a period of high eccentricity of planet c,
resulting in a particularly high maximum value of $\psi_{*b}$. The importance of this difference can be understood as follows.
In each model, the (average) value of $\psi_{*c}$ depends on its value before the scatter, as well as the change
in the inclination of c's orbit relative to the original invariable plane.
Since the orbit of c effectively coincides with the invariable plane once d has left the system, $i_b\simeq \psi_{*c}$ 
for a system in which there is no close encounter between b and c (at least for
the cases considered here for which $\psi_{*b}(0)=0$), and the variation in $\psi_{*b}$ is approximately equal
to $2\psi_{*c}$.
On the other hand, $i_b$ is significantly greater than $\psi_{*c}$ in model 3 because of the close encounter
of b and c, and as such the maximum value of $\psi_{*b}$ is high.

\begin{figure}
\centering
\includegraphics[width=170mm]{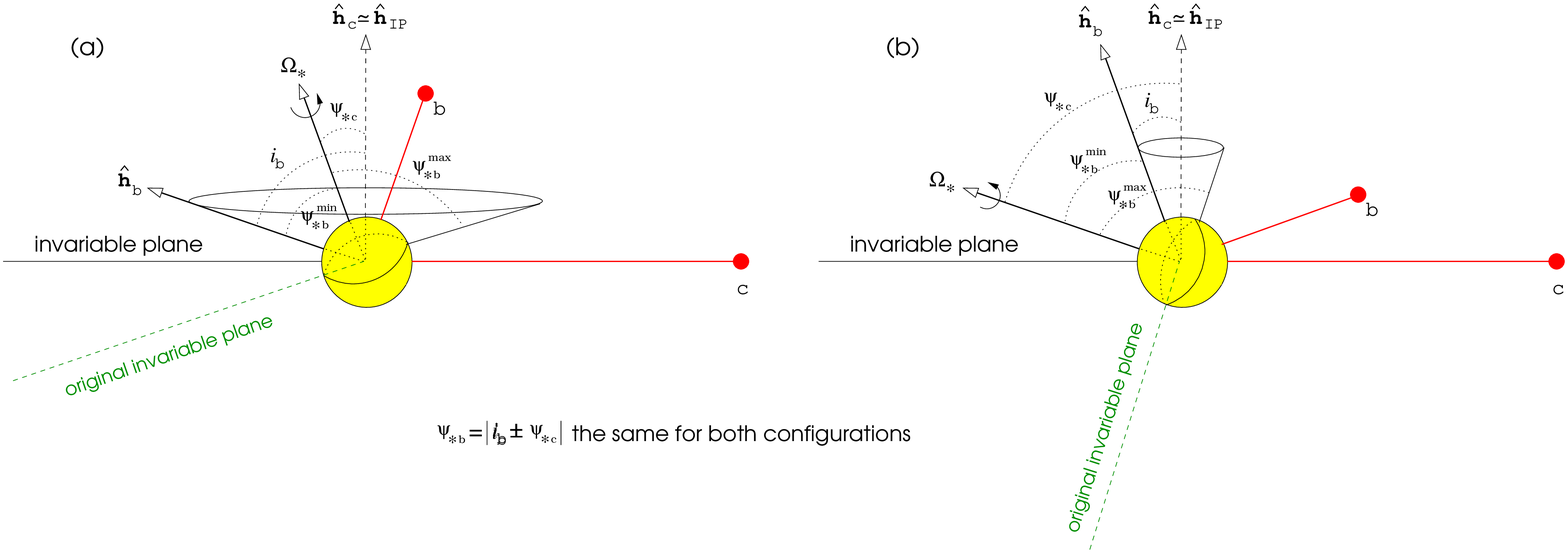}
\caption{Illustration of high stellar obliquity in cases where (a) planet c is scattered through a small angle 
during its interaction with a third planet, and planet b is scattered through a {\it large} angle 
during its interaction with planet c, 
and (b) planet c is scattered through a large angle 
during its interaction with a third planet (or passing star), and planet b is also scattered through a large angle 
during its interaction with planet c, but in such a way that it ends up with a small value of $i_b$.
Here we are assuming that the stellar spin direction is parallel to the invariable plane normal {\it before} 
interaction with the fourth body, and that the system was originally coplanar.
While both configurations have the same range of values of $\psi_{*b}=|i_b\pm\psi_{*c}|$,
configuration (a) is ruled out for HAT-P-13 because $54<i_b<126^\dg$ (see Section~\ref{high}), while
configuration (b) cannot be ruled out (although it would probably
require a close encounter with a passing star rather than the escape of a companion planet
to cause such a dramatic change in c's orbit). 
}
\label{obliquity}
\end{figure}

Figure~\ref{obliquity}
illustrates the mechanics of stellar obliquity in a two-planet system for the cases (a)
$\psi_{*c}=20^\dg$ and $i_b=70^\dg$ (a configuration ruled out for the HAT-P-13 system
because $54<i_b<126^\dg$; see Section~\ref{high})
and (b) $\psi_{*c}=70^\dg$ and $i_b=20^\dg$,
so that $50^\dg\leq \psi_{*b}\leq 90^\dg$ for both. Also indicated
is the original invariable plane in the case that the system contained a third planet which has since escaped,
and whose angular
momentum was such that the stellar spin axis was parallel to the invariable plane normal.
The ramifications for the origin of retrograde systems are clear (although configuration (b)
would probably
require a close encounter with a passing star rather than the escape of a companion planet
to cause such a dramatic change in c's orbit). 
Given the amount of angular momentum transferred from c to b depends on all the system parameters,
it is easily conceivable that a scattering scenario similar to the one described here (either bound or flyby,
perhaps involving exchange of c and d) could produce retrograde systems such as HAT-P-7 and WASP-17.
It is interesting to note that it is possible for the relative inclination of a system to pass through $90^\dg$
without destroying planet b through Kozai oscillations as long as $\gamma$ is large enough (see Section~\ref{high}).

We end by considering the effect on the mutual inclination of non-zero stellar obliquity with respect to the invariable plane,
which can be significant if $\theta_*$ is significantly non-zero.

\subsubsection{Effect on the mutual inclination of
non-zero stellar obliquity with respect to the invariable plane}\label{nonzeroobliquity}

\begin{figure}
\centering
\includegraphics[width=120mm]{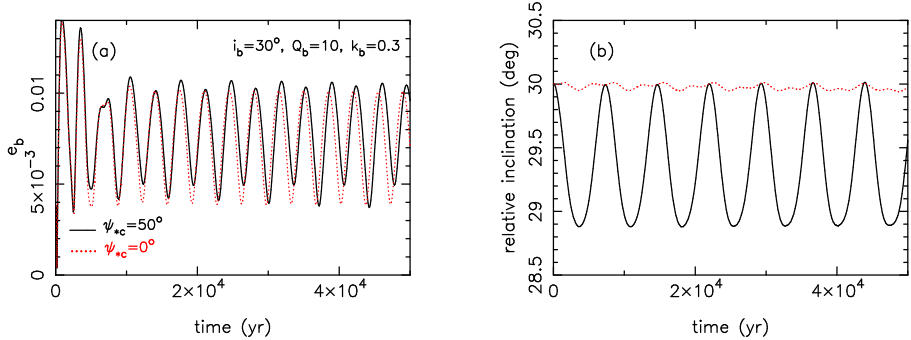}
\caption{The effect of non-zero stellar obliquity {\it relative to the invariable plane}
on the relaxed state for (a) $e_b$ and (b) $i_b$. The limit-cycle amplitude (ie, the amplitude of variation of $e_b$)
varies regularly on a timescale of 
twice the limit-cycle
period, the latter being $2\pi/2\dot\omega_b$, and its modulation period is slightly longer.
The amplitude of variation of $i_b$ is significantly enhanced by an amount proportional to $\sin(2\theta_*)$.
For comparison, the variation of $i_b$ for the case $\theta_*(0)=0$ is shown (red curve in panel (b)).
The small variation is due to nutation of the stellar spin axis through about $2.5^\dg$.
}
\label{obliquity-comp}
\end{figure}

The numerical results presented in Section 3 assume zero stellar obliquity relative to
the invariable plane, that is, $\theta_*=0$. 
Consequences of this are that the relaxed state illustrated in Figure~\ref{limit-cycle} (red curves) and Figure~\ref{eeq4}(a)
exhibits constant amplitude variations, and that the inclination $i_b$ is effectively constant. 
The latter is clearly not the case in the three models discussed above (panel (c) of Figures~\ref{model1},
\ref{model2} and \ref{model3}).
Figure~\ref{obliquity-comp}(a)
shows how the amplitudes of variation of $e_b$ and $i_b$ are modulated when $\theta_*$ is non-zero. Here 
$\theta_*=50^\dg$, $i_b=30^\dg$ and $Q_b=10$, and the initial values of $e_b$ and $\eta$
are $e_b^{(av)}$ and zero respectively. The limit-cycle amplitude varies noticeably and the modulation period is slightly longer,
with the behaviour persisting during a $10^6$ year integration.
This can be understood as follows. The torque on the orbit of planet b due to the star's spin oblateness is given by
equation (48) in \citet{ML}, and this contributes to the rate of change of $i_b$ according to equation (29) of the same paper.
Expressing the stellar spin vector $\bOmega_*$ and the 
basis vectors $\hat{\bf e}_b$, $\hat{\bf q}_b$ and $\hat{\bf h}_b$ referred to in \citet{ML}\footnote{See also Section~\ref{seceReq} here.}
in terms of the invariable plane reference basis via 
$\theta_*$, the star's node angle $\varphi_*$,
and the Euler angles $\omega_b$, $\Omega_b$ and $i_b$,
the contribution to $di_b/dt$ from the star's spin oblateness becomes
\be
\left.\frac{di_b}{dt}\right |_*=-\frac{1}{2}\nu_bk_*\left(\frac{R_*}{a_b}\right)^5\left(\frac{\Omega_*}{n_b}\right)^2
\left[\cos\theta_*\cos i_b\cos(\Omega_b-\varphi_*)+\ff{1}{2}\sin\theta_*\sin i_b\sin(2\Omega_b-2\varphi_*)\right].
\label{didtstar}
\ee
Taking $\theta_*$, $i_b$ and $\varphi_*$ to be approximately constant over an orbit precession cycle
(the precession rate of the stellar spin axis is 40 times slower than the orbit precession rate for this example),
and putting $\Omega_b=\dot\Omega_b t+\Omega_b(0)$
where $\dot\Omega_b$ is given by \rn{Ob} and is also approximately constant, 
the variation in $i_b$ due to the star's spin oblateness, $ \Delta i_b$, 
is obtained by integrating \rn{didtstar} over an orbit precession cycle to give
\be
\Delta i_b=\gamma_*^{spin}\varepsilon_c^3\sin (2\psi_{*c})
\ee
where
$\gamma_*^{spin}$ is the contribution to $\gamma$ from the star's spin quadrupole moment and is given in
Table~\ref{planet} for HAT-P-13 for $m_c=m_c^{min}$. 
Putting $\psi_{*c}=50^\dg$ then gives $\Delta i_b=1.1^\dg$, in good agreement
with Figure~\ref{obliquity}(b). Such a variation in turn produces a variation in $e_b^{max,min}$ of the order seen in panel (a)
of the same figure.

\section{Summary}\label{discussion}

The ideas and results presented in this paper can be summarized as follows:

\sspace 1. Generalizing the results of \citet{puffball} to non-coplanar systems
for which most of the angular momentum of the system resides in the outer orbit,
we find that under the action
of tidal dissipation in planet b, the system evolves to a limit cycle rather than a fixed point
in $e_b-\eta$ space, with the average value of $e_b$, $e_b^{(av)}$, decreasing and the limit cycle
amplitude increasing with increasing mutual inclination, and
$\lim_{i_b\rightarrow 0}e_b^{(av)}=e_b^{(eq)}$. For the HAT-P-13 system, limit cycle
behaviour occurs for $i_b\lapp 33^\dg$ (libration of $\eta$ around $2n\pi$)
and $46^\dg\lapp i_b\lapp 54^\dg$  (libration of $\eta$ around $(2n+1)\pi$).
No limit cycle exists for $33^\dg\lapp i_b\lapp 46^\dg$ ($\eta$ circulates).

\sspace 2. For systems with $54^\dg<i_b<90^\dg$ ($90^\dg<i_b<126^\dg$),
Kozai oscillations coupled with tidal dissipation in planet b act to reduce (increase) the mutual
inclination until $i_b<54^\dg$ ($i_b>126^\dg$)
on a timescale much less than the age of the system. We conclude that the HAT-P-13 system
cannot have a mutual inclination between $54^\dg$ and $126^\dg$ for $Q_b\lapp 10^6$.

\sspace 3. For retrograde systems, the limit-cycle behaviour is the mirror image of that for prograde systems,
apart from a slightly different limit-cycle frequency (compare \rn{Wwr} with \rn{Ww}).

\sspace 4. The analysis and conclusions of \citet{batygin} for the HAT-P-13 system 
are valid as long as the mutual inclination of planets b and c
is less than around $10^\dg$ (which may well be the case; see next point). 
For higher prograde values of $i_b$, a measurement of $e_b$ does not unambiguously
determine $k_b$, although one can make arguments about the likelihood of a system being near
the top or bottom of the modulation cycle of $e_b$. Similar statements hold for mutually retrograde systems.

\sspace 5. We have derived a relationship between the average eccentricity, $e_b^{(av)}$,
the equilibrium radius of planet b, $R_b^{(eq)}$, its $Q$-value and its core mass (Figure~\ref{eReq}), and conclude that
for $Q$-values greater than the lower bound imposed by the timescale for decay of $e_b^{(av)}$, 
the orbits of planets b and c
are likely to be either near prograde coplanar,
or have mutual inclinations between around $130^\dg$ and $135^\dg$.
Lower rather than higher core masses are favoured.
More accurate measurements of $e_b$ and $R_b$ will allow refinement of these statements.

\sspace 6. Inclined systems cannot relax to the coplanar prograde or retrograde state as long as $e_b^{(av)}>0$,
and instead relax to a mutual inclination given by one of the roots of $\Psi(i_b)=0$ or $\Psi(\pi-i_b)=0$ (see equation \rn{Psi}).
This will occur as long as $\tau_i<\tau_c$ and $\tau_i<\tau_a$, both true for the HAT-P-13 system,
however, in this case, $\tau_i$ is much greater than the age of the system.

\sspace 7. A viable formation scenario for the HAT-P-13 system is that it originally contained
a third planet which was scattered out of the system 
when the protoplanetary disk density dropped
below the critical level for stability. Such a scenario is capable of producing a high eccentricity for planet c
as long as the mass of planet c is sufficiently high, and may also produce significant mutual inclination,
stellar obliquity and inflated planetary radii.

\sspace 8. A Rossiter-McLaughlin measurement of the sky-projected stellar obliquity relative to planet b, 
together with a measurement of the mutual inclination, will allow us to constrain the stellar obliquity relative
to planet c and hence obtain knowledge about the formation history.

\vspace{3mm}

In conclusion, it is likely that many more HAT-P-13-like systems will be discovered in the future including systems
in which the outer body is a binary star companion. Such systems will contribute a wealth of information 
not only about the internal structure of the short-period planet, but also about the formation history of the system. 

\sspace {\it Note added in proof}:\\
Since this paper was submitted more refined data for the HAT-P-13 system have
become available \citep{winn10}, in particular, a new estimate for the eccentricity of planet b
of $0.0142^{+0.0052}_{-0.0044}$.
Moreover, there is now evidence for a distant third body
in the system, as well as a Rossiter-McLaughlin measurement of the sky-projected stellar obliquity, 
the latter strongly suggesting that the stellar spin and the orbit normal of planet b are
aligned. The conclusions drawn here remain valid, in particular those
regarding coplanarity or otherwise when the refined
value of $\omega_b-\omega_c$ is used.

\section*{Acknowledgments}

The author wishes to thank Dan Fabrycky for inspirational discussions and encouragement, 
for carefully reading the manuscript,
and especially for his generosity in allowing her to include his elegant argument ruling out
the 45 - $50^\dg$ and retrograde coplanar configurations for the HAT-P-13 system. Thanks
also go to Eric Ford for a very valuable comment.

\appendix
\section{Secular equations for a Newtonian point-mass system}
In \citet{secular}, the secular equations governing the evolution of the orbital
elements of the inner and outer binaries of a hierarchical triple are given to arbitrary order.
These are derived using a spherical harmonic expansion of the disturbing function ${\cal R}$ expressed
in Jacobi coordinates, the latter having the dimensions of energy and defined to be such that the total energy is
\be
E=-\frac{1}{2}\frac{Gm_*m_b}{a_b}-\frac{1}{2}\frac{G(m_*+m_b)m_c}{a_c}-{\cal R}.
\ee
Noting from the numerical solution for the HAT-P-13 system
presented in Section~\ref{inclined} that $e_b$, $a_b/a_c$ and ${\rm max}(\sin i_c)$ 
are all of order 0.01, and 
taking the invariable plane to be the reference plane,
the orbit-averaged disturbing function to order sufficient in $e_b$, $a_b/a_c$ and $\sin i_c$ to produce the
dominant terms of each of the rates of change of the elements is
\be
\tilde{\cal R}=\tilde{\cal R}_q+\tilde{\cal R}_o,
\label{disturbing}
\ee
where the quadrupole and octopole contributions are given by
\bea
\tilde{\cal R}_q&=&\ff{1}{4}\mu_b a_b^2 n_b^2\left(\frac{m_c}{m_*}\right)\left(\frac{a_b}{a_c}\right)^3
\varepsilon_c^{-3}
\left\{(1+\ff{3}{2}e_b^2)f_1(i_b)f_1(i_c)+\ff{15}{4} e_b^2 \,\sin^2 i_b\cos (2\omega_b)\right.\next
&&\left.\hspace{4cm} 
+\ff{3}{4}\cos(\Omega_b-\Omega_c)\sin 2i_b\sin 2i_c
+\ff{3}{4}\cos(2\Omega_b-2\Omega_c)\sin^2i_b\sin^2i_c
\right\}
+{\cal O}(x^5)
\label{Rq}
\eea
and
\be
\tilde{\cal R}_o=-\ff{15}{16}\mu_b a_b^2 n_b^2\left(\frac{m_c}{m_*}\right)\left(\frac{a_b}{a_c}\right)^4
e_b e_c\varepsilon_c^{-5}\left\{f_2(i_b)\cos(\varpi_b-\varpi_c)
+g_2(i_b)\cos\left(\varpi_b+\varpi_c-2\Omega_b\right)\right\}+{\cal O}(x^7),
\ee
respectively, where
$x$ represents any one of $e_b$, $a_b/a_c$ or $\sin i_c$ (for example,
$x^2$ may represent $e_b\sin i_c$, and $\cos i_c=1+{\cal O}(x^2)$).
Note that here we have assumed that $m_b/m_*\ll 1$ and $m_c/m_*\ll 1$ (in contrast, the 
equations in \citet{secular} are completely general). 
The inclination functions $f_1(i_b)$, $f_2(i_b)$ and $g_2(i_b)$
are listed as follows together with $f_3(i_b)$, $f_4(i_b)$ and $h_4(i_b)$ which appear in the evolution equations below,
and $h_3(i_b)$ which appears in the theory of relaxed retrograde orbits (Section~\ref{retrograd}):
\bea
f_1(i_b)&=&\ff{1}{2}(3\cos^2i_b-1),\label{A5}\\
f_2(i_b)&=&\ff{1}{8}(1+\cos i_b)(15\cos^2i_b-10\cos i_b-1) \label{A6}\\
f_3(i_b)&=&\ff{1}{2}(5\cos^2i_b-2\cos i_b-1) \label{A7}\\
f_4(i_b)&=&\ff{1}{4}(15\cos^2i_b-10\cos i_b-1) \label{A8}\\
g_2(i_b)&=&\ff{1}{8}(1-\cos i_b)(15\cos^2i_b+10\cos i_b-1) \label{A10}\\
h_3(i_b)&=&\ff{1}{2}(5\cos^2i_b+2\cos i_b-1) \label{A7b}\\
h_4(i_b)&=&\ff{1}{4}(15\cos^2i_b+10\cos i_b-1) \label{A9}
\eea
The inclination functions \rn{A5}-\rn{A10} have the symmetry properties
$f_1(i_b)=f_1(\pi-i_b)$,
$f_2(i_b)=g_2(\pi-i_b)$,
$f_3(i_b)=h_3(\pi-i_b)$,
$f_4(i_b)=h_4(\pi-i_b)$,
$f_n(0)=1$, 
$g_2(\pi)=h_3(\pi)=h_4(\pi)=1$
and $g_2(0)=f_2(\pi)=0$,
all of which are relevant for the comparison of prograde and retrograde systems
(see Section~\ref{retrograd}).

Given our aim of understanding the long-term behaviour of HAT-P-13-like systems and
having demonstrated empirically that they maintain small values of 
$\sin i_c$ and evolve towards small values of $e_b$ 
on a timescale equal to three times the tidal circularization timescale for {\it any} initial relative inclination,
we follow \citet{puffball} and
write down the equations governing the secular evolution of the orbital elements,
retaining only leading order terms in $e_b$, $a_b/a_c$ and $\sin i_c$.
Using \rn{disturbing}, Lagrange's planetary equations are \citep{MD}\footnote{Note that since our $\tilde{\cal R}$
has the dimensions of energy, the usual equations are divided by $\mu_b$ and $\mu_c$ for the rates
of change of the inner and outer orbital elements respectively.}
\be
\frac{de_b}{dt}=\ff{15}{8}C_b^{(q)} e_b \sin^2 i_b\sin (2\omega_b)
-\ff{15}{16}C_b^{(o)} e_c\left[f_2(i_b)\sin(\varpi_b-\varpi_c)+g_2(i_b)\sin(\varpi_b+\varpi_c-2\Omega_b)\right]
+{\cal O}(x^5),
\label{ep}
\ee
\be
\frac{di_b}{dt}=\ff{3}{8}C_b^{(q)}
\sin(\Omega_b-\Omega_c)\cos i_b\sin 2i_c
-\ff{15}{32}C_b^{(o)}e_b e_c\sin(2 i_b)\left[ f_4(i_b)\sin(\varpi_b-\varpi_c)
-h_4(i_b)\sin(\varpi_b+\varpi_c-2\Omega_b)\right]
+{\cal O}(x^5),
\label{ip}
\ee
\be
\frac{d\varpi_b}{dt}=\ff{3}{4}C_b^{(q)}
\left[f_3(i_b)+\ff{5}{2}\sin^2 i_b\cos (2\omega_b)\right]
-\ff{15}{16}C_b^{(o)}\left(\frac{e_c}{e_b}\right)\left[f_2(i_b)\cos(\varpi_b-\varpi_c)+g_2(i_b)\cos(\varpi_b+\varpi_c-2\Omega_b)\right]
+{\cal O}(x^4),
\label{wp}
\ee
\be
\frac{d\Omega_b}{dt}=-\ff{3}{4}C_b^{(q)}\cos i_b+{\cal O}(x^4),
\label{Ob}
\ee
\be
\frac{d e_c}{dt}=\ff{15}{16}C_c^{(o)}e_b\left[f_2(i_b)\sin(\varpi_b-\varpi_c)
-g_2(i_b)\sin(\varpi_b+\varpi_c-2\Omega_b)\right]+{\cal O}(x^{13/2}),
\label{ec}
\ee
\be
\frac{di_c}{dt}=-\ff{3}{8}C_c^{(q)}\varepsilon_c^{-1} \sin(\Omega_b-\Omega_c)\sin 2i_b\cos i_c+{\cal O}(x^{11/2}),
\ee
\be
\frac{d\varpi_c}{dt}=\ff{3}{4}C_c^{(q)}f_1(i_b)+{\cal O}(x^{9/2})
\label{wc}
\ee
and
\be
\frac{d\Omega_c}{dt}=-\ff{3}{8}C_c^{(q)}\varepsilon_c^{-1}\cos(\Omega_b-\Omega_c)\sin 2i_b
\frac{\cos 2i_c}{\sin i_c}+{\cal O}(x^{7/2}),
\label{Oc}
\ee
where 
\be
C_b^{(q)}=n_b\left(\frac{m_c}{m_*}\right)\left(\frac{a_b}{a_c}\right)^3 \varepsilon_c^{-3},
\hh
C_b^{(o)}=n_b\left(\frac{m_c}{m_*}\right)\left(\frac{a_b}{a_c}\right)^4 \varepsilon_c^{-5}
=\left(\frac{a_b}{a_c}\right) \varepsilon_c^{-2}\,C_b^{(q)},
\ee
\be
C_c^{(q)}=n_c\left(\frac{m_b}{m_*}\right)\left(\frac{a_b}{a_c}\right)^2 \varepsilon_c^{-4}=
\left(\frac{m_b}{m_c}\right)\sqrt{\frac{a_b}{a_c}} \varepsilon_c^{-1}\,C_b^{(q)}
\hand
C_c^{(o)}=n_c\left(\frac{m_b}{m_*}\right)\left(\frac{a_b}{a_c}\right)^3 \varepsilon_c^{-6}.
\ee
In addition,
\be
\dot\omega_b=\dot\varpi_b-\dot\Omega_b.
\label{omega}
\ee
Our decision to retain or ignore each particular term is guided by numerical solutions for the relaxed state
of systems with arbitrary relative inclinations. For example, we have omitted the quadrupole term
$-\ff{15}{16}C_b^{(q)}e_b^2\sin(2 i_b)\sin(2\omega_b)$ in $di_b/dt$ which is responsible for Kozai oscillations
and has a modulation frequency of $2\dot\omega_b$; in the relaxed state this term is ${\cal O}(x^5)$ and
contributes negligibly to the dynamics, whereas the quadrupole term we {\it have} included dominates the 
behaviour of $i_b$ with its modulation frequency of $\dot\omega_b$. Moreover, 
while the octopole term is ${\cal O}(x^5)$, we have included it because it provides a timescale
for the slow decay of the mutual inclination (see Section~\ref{incdec}).

\label{lastpage}

\end{document}